\documentclass{aa}

\usepackage{graphicx}
\usepackage{txfonts}
\usepackage{natbib}
\usepackage{hyperref}
\usepackage[svgnames]{xcolor}
\usepackage{multirow}
\usepackage{bm}
\usepackage[normalem]{ulem}
\usepackage{threeparttable}
\usepackage{longtable}
\usepackage{longtable,lscape}
\usepackage[outercaption]{sidecap}

\newcommand\teff{$T_{\fontsize{6}{6}\selectfont \mbox{eff}}$}\normalfont
\newcommand\logg{$\log g$}

\newcommand\kmsec{km\,s$^{-1}$}
\newcommand{\kms}{km\,s$^{-1}$}
\newcommand{\vsini}{$v\sin i$}
\newcommand{\feh}{[Fe/H]}

\newcommand\Space{SP\_Ace}

\newcommand{\rotfit}{{\sf ROTFIT}}
\newcommand{\gaia}{{\it Gaia}}

\begin{document}

   \title{The open cluster NGC 2509. Stellar rotation and  main sequence turnoff extension from FLAMES spectroscopy}

   \author{C. Boeche\inst{1} 
          \and J. Alonso-Santiago\inst{2}
          \and A. Bragaglia\inst{3} 
          \and A. Frasca\inst{2} 
          \and A. Vallenari\inst{1} 
          \and I.N. Kallimanis\inst{1,4}
          \and R. Carrera\inst{3} 
          \and D. Bossini\inst{4}
          \and S. Lucatello\inst{1}
          \and V. D'Orazi\inst{5,6}
          \and G. Costa \inst{4}
          }

   \institute{INAF - Osservatorio Astronomico di Padova, vicolo dell'Osservatorio 5, 35122 Padova, Italy\\
          \email{corrado.boeche@inaf.it}
          \and
         INAF--Osservatorio Astrofisico di Catania, via S. Sofia 78, 95123 Catania, Italy \\
                   \email{javier.alonso@inaf.it}
         \and
         INAF - Osservatorio di Astrofisica e Fisica dello Spazio, via P. Gobetti 93/3, 40129 Bologna, Italy\\
              \email{angela.bragaglia@inaf.it}
        \and
          Universit\`a di Padova, Dipartimento di Fisica e Astronomia G. Galilei, Padova, Italy\\
         \and
          Department of Physics, University of Rome Tor Vergata, via della Ricerca Scientifica 1, 00133, Rome, Italy \\
        \and
          INAF Osservatorio Astronomico di Roma - Via Frascati 33, 00040, Monte Porzio Catone, Italy
}
   \date{Received ; accepted}

  \abstract {NGC\,2509 is a distant ($\approx$\,2.5\,kpc) and little-studied
  open cluster located in the third Galactic quadrant.  It is a moderately
  old cluster, whose age has not yet been precisely determined.  The
  main-sequence stars in NGC\,2509 follow a narrow distribution in the
  color-magnitude diagram, unlike other clusters of similar age.  In
  addition, its chemical composition has never been investigated.  To
  address these issues and characterize the cluster we performed moderate-
  and high-resolution spectroscopy with FLAMES@VLT of 132 stars, both dwarfs
  and giants, which represents a significant fraction ($\approx$\,73\,\%) of
  likely members.  We provide atmospheric stellar parameters and, for the
  first time, chemical abundances for 21 species with atomic numbers up to
  60.  In our analysis we followed two different methodologies, both of
  which will be used for the incoming WEAVE stellar surveys.  We find an
  average radial velocity for NGC\,2509 of 58.6$\pm$1.3\,km\,s$^{-1}$ and a
  mild supersolar metallicity ([Fe/H]$\approx$\,0.1\,dex).  This value is
  slightly higher than expected according to its galactocentric distance,
  but still compatible with the Galactic gradient.  From the lithium content
  of the dwarfs and the isochrone-fitting method we obtain an age for
  NGC\,2509 of 1.26$\pm0.3$Gyr.  The reddening across the cluster field is
  negligible ($A_V=0.25\pm0.02$\,mag).  The cluster peculiar main sequence
  turnoff is due to a narrow distribution of the rotational velocities
  peaking at \vsini\ $\approx80$~\kmsec, with little dispersion.  The
  chemical pattern of NGC\,2509 follows the Galactic trends shown by other
  open clusters in the Galactic thin disk.}

   \keywords{Open clusters and associations: general -- 
   Open clusters and associations: individual (NGC~2509) -- Stars: fundamental parameters -- Stars: abundances -- Stars: evolution  -- Galaxy: disk }

\titlerunning{The open cluster NGC~2509}

\maketitle
 
\section{Introduction}\label{sec_intro}
Open clusters (OC) have been extensively studied because they are both very
good tracers of the Galactic disk properties and the best test for
theoretical models.  For example, the public $Gaia$-ESO spectroscopic
survey (GES) observed more than 60 OCs
\citep[see][for details]{randich2022}, leading, among other significant
results, to a redetermination of the disk radial gradient
\citep{Magrini2023}, and to a comparison of Li abundances with stellar
models \citep{Magrini2021}. 
A number of programs involve high-resolution spectroscopy of OC stars.  We
cite, among others, the Open Cluster Chemical Abundances from Spanish 
Observatories survey \citep[OCCASO][]{2024A&A...687A.239C}, the Open Cluster 
Chemical Abundances and Mapping survey
\citep[OCCAM][]{otto26}, the Galactic Archaeology with HERMES survey
\citep[GALAH][]{kos2025}.

However, NGC~2509 is not included in these programs, and nor, in particular, in the
$Gaia$-ESO sample \citep{bragaglia2022}. We decided to amend the
situation given the interesting properties of this system.
\citet{carraro2007} presented %$V,V-I$ 
$VI$ photometry of this cluster and the derived age=1.2\,Gyr, $E(V-I)$=0.08\,mag, and
distance modulus 12.50\,mag, corresponding to a distance of 2.9\,kpc (see their
paper for a short discussion of inconsistencies in previous literature). 
\citet{cantat2020} and \citet{HR2024}, based on data from $Gaia$-DR2 and
DR3, respectively, derived very similar ages and distances (about 1.5\,Gyr
and 2.5\,kpc), while more discrepant values were found for reddening
($A_V$=0.118 and 0.23\,mag, respectively).  To our knowledge, there are only
two published determinations of metallicity, based on HARPS-N very
high-resolution spectra of a single star: \citet{Zhang21} measured
[Fe/H]=--0.1\,dex (adopting a local thermodynamic equilibrium (LTE)
analysis), while \citet{marina25} found [Fe/H]=0.21\,dex using non-LTE.

The most recent work dedicated to NGC~2509 was by \citet{deJuanOvelar2020},
who derived a younger age of 860\,Myr, a distance modulus of 12.36\,mag,
absorption A$_V$=0.75\,mag.  Their analysis of the photometry suggests solar
metallicity.
They studied this cluster in the framework of the extended main sequence (MS) turnoffs
(eMSTO), which are routinely observed in OCs similar in age to NGC~2509
\citep[see, e.g.,][and references therein]{cordoni24}, but, interestingly,
not in this cluster.  Indeed,
the eMSTO is an ubiquitous feature in intermediate age stellar clusters. 
It is usually attributed to the combined effect of changes in the internal
chemical and hydrostatic equilibrium structure due to stellar rotation
\citep{BastianMink2009} or to the effect of a prolonged star formation \citep[see][for instance]{2025A&A...701A.221S}. The fact that NGC~2509 presents a very narrow MS and MSTO
was interpreted as the effect of a peculiar rotation velocity distribution
among its stars by \citet{deJuanOvelar2020}, who proposed 
that the cluster has a narrow rotational velocity distribution and that most
of the MSTO stars rotate at about one-half the critical velocity
\citep[which, for stars of about 1\,Gyr, is on the order of about
300\,km~s$^{-1}$;][]{2009pfer.book.....M}.  However, this conclusion is
based on a comparison of the color-magnitude diagram (CMD) with stellar
isochrones, and no measurements of
rotation velocity are available in the literature.  The purpose of this paper
is to derive the properties of NGC~2509 from high-resolution spectroscopy to
clarify the nature of its peculiar CMD.

The paper is structured as follows.  It begins with the presentation of our
observations in Sect.~\ref{obs} and continues in Sect.~\ref{sec_analysis}
with the description of the dual approach performed in the spectral
analysis.  Specific details regarding the determination of the radial and
rotational velocities are set out in Sects.~\ref{sec_RV} and
~\ref{sec_vsini}, respectively.  In Sect.~\ref{sec_APs} we discuss the
findings obtained in this work, first comparing the results derived with
each methodology and then comparing these with the literature.  The study of
the age and the reddening across the cluster field is presented in
Sect.~\ref{sec_age} while in Sect.~\ref{discussion} we put the metallicity
and the chemical composition determined for NGC~2509 in this work in the
Galactic framework.  Finally, the paper concludes in Sect.~\ref{summary}
with a summary of our results.

\section{Observations} \label{obs}

We conducted our study by performing spectroscopy with the Fiber Large Array
Multi-Element Spectrograph \citep[FLAMES;][]{pasquini2002}, which is
mounted on the 8.2-m VLT UT2 telescope at ESO's Paranal Observatory in
Chile.  FLAMES feeds two different spectrographs, GIRAFFE and the 
Ultraviolet and Visual Echelle Spectrograph (UVES), which
cover the whole optical spectral range.  
The GIRAFFE spectrograph allows us to observe up to 132 objects, 
and the high-resolution gratings (HR) permit a resolving power, $R$, ranging
from  about 12\,000 to 37\,000, depending on the setup. 
On the other hand, UVES observes at most
eight objects, but with a higher resolution  ($R$=47\,000).
We selected our targets among the list of members reported in
\citet{HR2023}, who identified them by taking advantage of the $Gaia$-DR3
data.  
\citet{HR2024} identified 447 stars around the cluster center, 180 of which have a membership probability larger than 0.7 (and 167 of them have $G_{mag}\le 17$\,mag).
We kept only candidate members falling within the apparent diameter of FLAMES,
that is, 12.5\,arcmin and with $G\le17$\,mag.  
We excluded from the pool of
possible targets a few stars already available in archives, those with low
membership probability \citep[as determined by][]{HR2023,HR2024}, and those
having warnings by $Gaia$ (for instance, for duplicity).  We used the Fiber
Positioner Observation Support Software (FPOSS) tool to produce three observing blocks
(OB), trying to maximize the number of stars observed with UVES while keeping the
same stars for the GIRAFFE fibers as far as possible.  Table~\ref{t:list}
shows the list of stars observed. We put the UVES fibers mostly on giants, in
particular on red clump (RC) stars; however, we also targeted a few MS stars close to the MSTO.  GIRAFFE fibers were used for
red giant branch (RGB) and MS stars; in particular, we chose several stars
lying on the equal-mass binary sequence, clearly visible in the cluster CMD.

The observations were done using the U580 setup for UVES (covering the
wavelength range 4800--6800\,\AA) and the three high-resolution GIRAFFE
setups HR11, HR12, and HR15N (covering the following spectral ranges
5356--5597, 5821--6146, and 6470--6790\,\AA\ and providing nominal
resolutions\footnote{\url{https://www.eso.org/sci/facilities/paranal/instruments/flames/inst/specs1.html}}
of 29\,500, 20\,250, and 19\,200, respectively).  The observations were
conducted in service mode in three exposures of 2775\,s each, carried
out from December 2023 to March 2024 (see Table~\ref{t:logobs}).  They were
deliberately taken several weeks apart to better look for possible binaries. 
The spectra were reduced
and calibrated following the standard procedures implemented in the ESO
instrument pipelines. In total, we collected 325 spectra with GIRAFFE and 20 with UVES, corresponding to 132 different stars (115 GIRAFFE + 17 UVES).
This sample represents about 73\%  of the candidate members.

\begin{figure}[ht!]
\centering
\includegraphics[width=9cm]{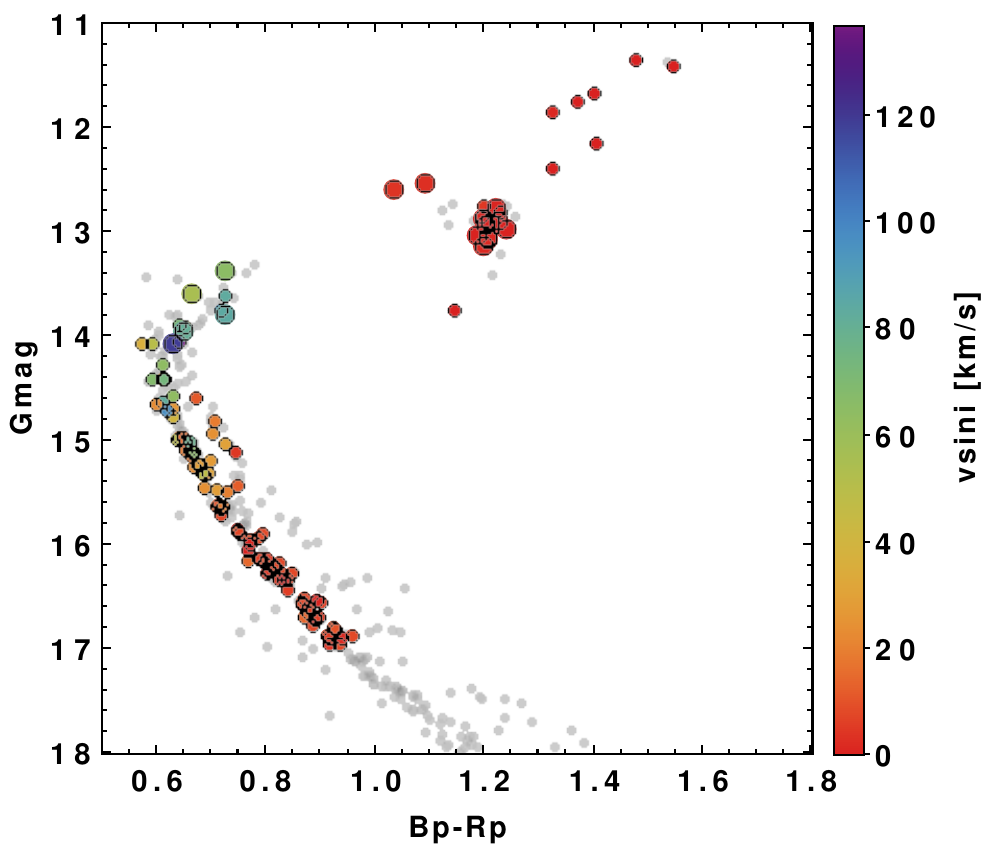}
\caption{Color-magnitude diagram of NGC~2509. The small gray points represent cluster members from \citet{HR2024}. The stars observed with UVES (big points) and GIRAFFE (small points) are colored as a function of \vsini\ computed as described in Sect.~\ref{sec_vsini}. }
\label{f:cmd_vsini}
\end{figure}

\begin{table*}[]
    \centering
    \caption{Log of the observations. }% \angie{o lo mettiamo come Tab. 1 o lo mettiamo come Tab A.1, qui non ha senso} }
    \begin{tabular}{ccclccccc}
\hline\hline
Observation & UT date and time  & exptime & Giraffe  & $R$ & airmass & seeing    & N.UVES  &N.GIR \\
 OB           & (at start)        & (s)     & setup    &            &         & (at start)& spectra &spectra \\
\hline
1of3 &2023-12-19T05:21:59 & 2775  & HR15N    &  15\,895   & 1.075 & 0.81 & 7  & 111 \\
2of3 &2024-01-23T03:56:01 & 2775  & HR11     &  21\,383   & 1.017 & 0.38 & 6  & 109 \\
3of3 &2024-03-09T02:53:26 & 2775  & HR12     &  16\,808   & 1.061 & 0.39 & 7  & 105 \\
\hline
    \end{tabular}

\tablefoot{For UVES, the U580 setup was used in all three cases. Some of the UVES and most of the GIRAFFE stars were observed in
more than one pointing and setup; see Table~\ref{t:list}.  The $R$ value
reported here is the observed one, computed as described in the text, lower
than the nominal values.} 
\label{t:logobs}
\end{table*}

\section{Data processing}
\label{sec_analysis}
Before analyzing the spectra, we
subtracted the sky from them.  For GIRAFFE, we took the median of the fibers
pointing the sky for each wavelength interval (HR11, HR12, and HR15N).  For UVES we subtracted the only sky fiber
available for that observing run for each wavelength interval (REDL and REDU).  Then, we normalized the spectra to the
continuum using an ad hoc routine made by us (see Appendix~\ref{sec_norm}).

We then performed spectroscopic analysis using two different approaches that
will be used for the analysis of incoming data that the survey WEAVE
\citep[WHT Enhanced Area Velocity Explorer; see][and references
therein]{Jin2024} will provide soon.  On the one hand, we used the \Space\
code that produces atmospheric parameters (APs) and elemental abundances. 
On the other hand, we used the combination of the codes \rotfit, to
calculate the APs, and SYNTHE, to carry out the chemical characterization. 
In the following, we outline the features of both methodologies.

\subsection{\Space\ }
We ran the software \Space\ \citep{Boeche2016,Boeche2018,Boeche2021} 
in order to derive stellar parameters and
chemical abundances from the GIRAFFE and UVES spectra. \Space\ relies on
a general curve of growth (GCOG) library where the equivalent widths (EWs) of the absorption
lines are stored as functions of the stellar parameters and chemical
abundances. \Space\ uses such GCOG functions to fabricate on-the-fly many template spectra searching through 
the stellar parameters and chemical abundances space for the one that best fits
the observed spectrum (i.e., that minimizes the $\chi^2$;  
see \citealt{Boeche2016} for details). In particular, we employed a
customized version of \Space\ developed for the WEAVE project based on the
most recent version of \Space\ \citep{Boeche2021}. This
version of \Space\ gets as input two separate wavelength windows and fits for
each of them distinct line profile widths, which allows it
to handle spectra that cover two windows with different resolutions at once.
The two windows must be located inside the extension of the \Space\
wavelength coverage (4800--6800\AA), they must not overlap, and their limits
are set by the user\footnote{In this work the two windows are approximately
5596--6137\AA\ and 6446--6816\AA\ for GIRAFFE; 4777--5796\AA\ and
5834--6825\AA\ for UVES.  The limits can change by a few angstroms for
different spectra.  \Space\ cuts the upper limit at 6800\AA.}.

Since \Space\ processes only normalized spectra at rest,
we first applied our normalization routine to each interval, 
and then corrected them for the radial velocity (RV).
Our normalization routine removes the cosmic rays and convolves 
the flux with a Gaussian kernel to normalize the flux to one.
Then we measured the RV of each interval using
the cross-correlation task {\it fxcor}  of the 
IRAF\footnote{IRAF is distributed by the National Optical Astronomy
Observatory, which is operated by the Association of the Universities for
Research in Astronomy, Inc.  (AURA), under cooperative agreement with the
National Science Foundation. We used the version V2.16.1.} package and the Sun
spectrum as a template\footnote{This procedure was applied for \Space\ only.} and corrected for it
with the IRAF task {\it dopcor}. Those RVs are of sufficient quality to be used for our purpose, i.e., spectrum correction. However, in the following, the cluster properties will be discussed using the higher precision \rotfit\ RV determinations.  After that, for each star we joined the
normalized RV corrected intervals into one single spectrum to be fed to
\Space.  The GIRAFFE and UVES spectra come in three and two
wavelength intervals, respectively. In the case of three wavelength intervals,
\Space\ treats the HR11 and HR12 intervals as one and HR15N as distinct and
estimates two different resolutions for these two parts. However, the stellar
parameters and chemical abundances are derived  from the  $\chi^2$ minimization, over all the intervals.  For stars that were observed twice, we derived the parameters and then
took the mean of the values. Some spectra were double lined or had high
\vsini\ (as discussed in Sect.~\ref{sec_rotfit}).
Such stars have been measured, but their results are dropped because
\Space\ cannot correctly handle double-lined or wide line profiles, and
this leads to unreliable results.
Thus, we selected as good quality results the ones derived from i) single-lined 
spectra, ii) spectra with $v \sin{i}$ < 10 \kmsec, and iii) spectra with
S/N$>$40 (the latter chosen from the \Space\ results). In total we analyzed 
98 GIRAFFE and 17 UVES stars for which we provide stellar parameters and chemical abundances for the elements Na, Mg, Al, Si,
Ca, Sc, Ti, V, Cr, Mn, Fe, Co, Ni, Cu, Y, Zr, Ba, La, Ce, and Nd (see Table~\ref{t:TGM_ABD_SPACE}).
Most of the GIRAFFE spectra have low S/N
while the UVES spectra all have high S/N\footnote{Note that \Space\ underestimates the S/N for high- resolution spectra; 
in this case the S/N of the UVES spectra are all underestimated.}.

\subsection{\rotfit\ and SYNTHE}\label{sec_rotfit}

We also used the code \rotfit\ \citep[e.g.,][]{Frasca2006} to measure
RV, \vsini, and atmospheric parameters, \teff, \logg, and \feh, with
the version of the code adopted for the GES
\citep[e.g.,][]{Frasca2015, Lanzafame2015}.  The template grid is the same
as that used by the OACT (Osservatorio Astrofisico di Catania) node within
the GES and is composed of high-resolution spectra of slowly rotating stars
(\vsini\,$\leq3$\,\kms) with a low activity level and known stellar
parameters.  These spectra were retrieved from the ELODIE archive
(R$\simeq$42,000; \citealt{Moultaka2004}). 
In addition \rotfit\ can use the synthetic BTSettl spectra.

The main steps of the analysis can be summarized as: {\rm i)}
re-normalization of the analyzed spectral segments by a fit of a low-order
polynomial; {\rm ii)} measure of the RV by the cross-correlation with the 
template spectrum that gives the highest peak (see
Sect.\,\ref{sec_RV}); {\rm iii)} determination of the APs and \vsini\ by
$\chi^2$ minimization of the residuals of the differences observed minus
templates, with each template brought to the target-spectrum
resolution and rotationally broadened by convolution with a
linear-limb-darkened rotational profile of varying \vsini; {\rm iv)}
spectral type (SpT) classification by taking that of the template with the
minimum $\chi^2$.  For details of the procedure, the reader is referred to
\citet{Frasca2015} and \citet{Lanzafame2015}.

For the APs derivation, the UVES spectra were split into segments of 100\,\AA\ each, which were
independently analyzed\footnote{The motivation for restricting the analysis to 100~\AA\ intervals stems from the
wavelength-dependence of the rotational kernel, which makes the simultaneous
modeling of multi-thousand angstrom ranges impractical.}.  The final parameters were obtained as the weighted
averages of those from individual segments, as described in detail in the
above mentioned papers.  As the three GIRAFFE setups have different
resolution and sampling, for each star we independently analyzed all the
spectra and compared the results.  In general, we find good agreement
between the APs derived with the three setups, with HR11 and HR12 showing
the best consistency with each other (see Fig.\,\ref{Fig:APs}).  Therefore,
we decided to take the weighted average (with variance-defined weights) of
the APs derived with HR11 and HR12 as the final parameters.  Excluding 
seven double-lined spectroscopic binaries (SB2; see Sect.~\ref{sec_RV}), for which no AP has been released, we have HR11+HR12 data for 93
sources.  For an additional eight stars, we only considered the HR11 results,
while for the other six we have only HR12; a single object has only an HR15N
spectrum.  In total, we provide APs, \vsini, and RV for 108 stars with
GIRAFFE spectra, plus the 17 stars observed with UVES. 
Our results are reported in Table~\ref{t:list}.

We also derived the chemical abundances of the giants observed with UVES, 12
stars in total.  As in previous works \citep[see, e.g.,  ][for further
details]{Radcliffe, M39} we resorted to the spectral synthesis technique. 
We first calculated the KURUCZ stellar models
\citep{Kurucz1993a,Kurucz1993b} for each of the stars in our sample from the
APs derived with \rotfit.  Then, the
corresponding synthetic spectra were generated using the {\sf SYNTHE}
code \citep{Kurucz1981}.  They were degraded to the UVES resolution and
broadened, taking into account both the instrumental and the rotational
profiles.  The observed spectra were subsequently compared with the
synthetic ones, deriving the chemical abundances of the former from the
$\chi^2$-minimization of the residuals.  We independently performed this
analysis in spectral segments of 50\,\AA\, obtaining the final abundances
for the star as the average of the values calculated in each segment, after
applying a 2-$\sigma$ clipping filter to avoid possible outliers.  Errors
represent the standard deviation among all values considered. We
investigated 20 elements with atomic numbers up to 60, namely Na, Mg, Al, Si,
Ca, Sc, Ti, V, Cr, Mn, Fe, Co, Ni, Zn, Y, Zr, Ba, La, Ce, and Nd (see Table~\ref{tab_abb_uves}).

\section{Radial velocity}
\label{sec_RV}

The RV values of the GIRAFFE spectra are derived by cross-correlating
each target spectrum with a list of templates selected from the grid adopted
by the code \rotfit.  Indeed, this is the first step of the \rotfit\ analysis
summarized in Sect.\,\ref{sec_rotfit}. 
To compute the cross-correlation function (CCF) for the HR15N setup, we have
excluded the H$\alpha$ line, whose wings would broaden the CCF peak. 
For RV measurement the
UVES spectra have been split into ten segments of 200\,\AA\ each from
4800\,\AA\, to 6800\,\AA, and the cross-correlation is carried out
separately in each region, after masking very broad lines, such as H$\beta$,
H$\alpha$, and \ion{Na}{i}\,D, as well as the regions strongly affected by
telluric absorption.  Synthetic BTSettl spectra \citep{allard2012}
downgraded to the spectral resolution of UVES and resampled at the spectral
points of the target spectra have been used as RV templates.

To measure the RV, for both GIRAFFE and UVES, we determined the centroid of
the CCF peak by fitting it with a Gaussian.  We used the
\textsc{curvefit} procedure \citep{Bevington}, taking into account the CCF noise,
$\sigma_{\rm CCF}$.  The latter was evaluated as the standard deviation of
the CCF values in two windows on the two sides of the peak.  The RV error
per each setup or spectral segment, $\sigma_{\rm RV}$, was estimated as the
error of the center of the Gaussian fitted to the CCF \citep[see
also][]{Frasca2019}.  For the UVES spectra, the RV and associated
uncertainty were calculated as the weighted mean and the weighted standard
deviation of the RVs obtained from individual segments, adopting the inverse
variance weight $w=1/\sigma_{\rm RV}^2$.

For seven stars, we noted, at least in one epoch, two distinct CCF peaks
that identified them as SB2s.  In
these cases, we evaluated the RV of the two components using a
two-Gaussian fit (see Fig.\,\ref{Fig:SB2} for some examples).  The radial
velocity of the primary (higher CCF peak) and secondary components of these
SB2s are reported in Table\,\ref{Tab:SB2}.  Moreover, for the other seven stars
with a single CCF peak, we found a variable RV, that is, $|\Delta RV|\geq
3\sqrt{\sum \sigma_{RV}^2}$, where $\sigma_{RV}$ is the RV error, in at
least two epochs.  We cite these objects along with their RV values in
Table\,\ref{Tab:SB1} and consider them as likely single-lined
spectroscopic binaries (SB1s).

\begin{figure}[htb]
\hspace{-.7cm}
\includegraphics[width=9.8cm]{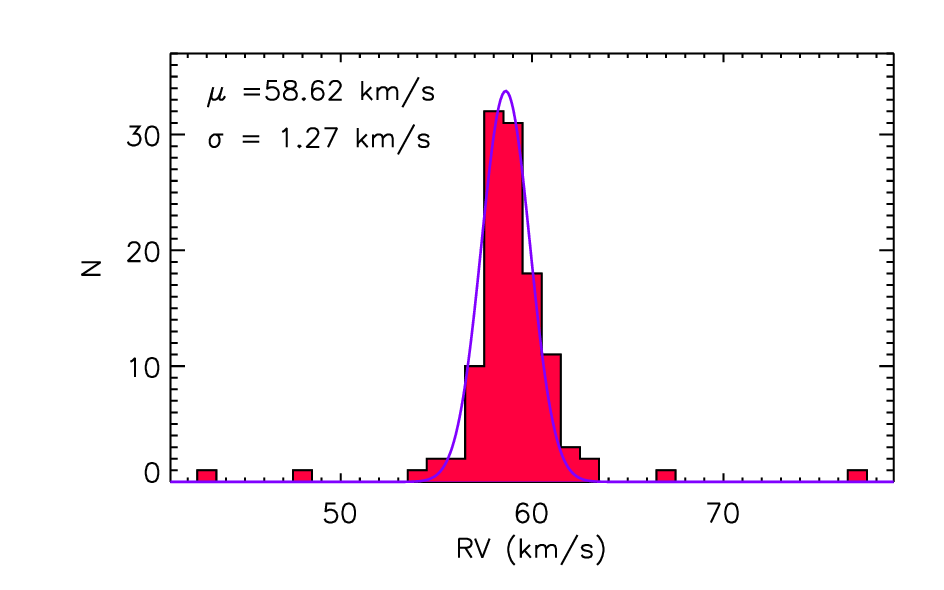}
\caption{RV distribution of the cluster members (red histogram).  The
Gaussian fit is overlaid with a full black line; the center ($\mu$) and
dispersion ($\sigma$) of the Gaussian are also marked.}
\label{Fig:RV_distr}
\end{figure}

With the exception of the seven SB2s, we have computed the weighted average
of the RV values obtained with the different setups and reported these
values in Table\,\ref{t:list} along with the atmospheric parameters and
\vsini\ measured with \rotfit.   The RV distribution is depicted in
Fig.\,\ref{Fig:RV_distr}.  With the exception of the few discrepant values
related to the likely SB1 binaries, the distribution displays a single
symmetric peak that is well fitted with a Gaussian, centered at 58.62\,\kms,
with a dispersion $\sigma=1.27$\,\kms.

\section{Rotation velocity and resolving power}\label{sec_vsini}

\begin{figure*}[th]  
\begin{center}
\hspace{-0.4cm}
\includegraphics[width=6.75cm]{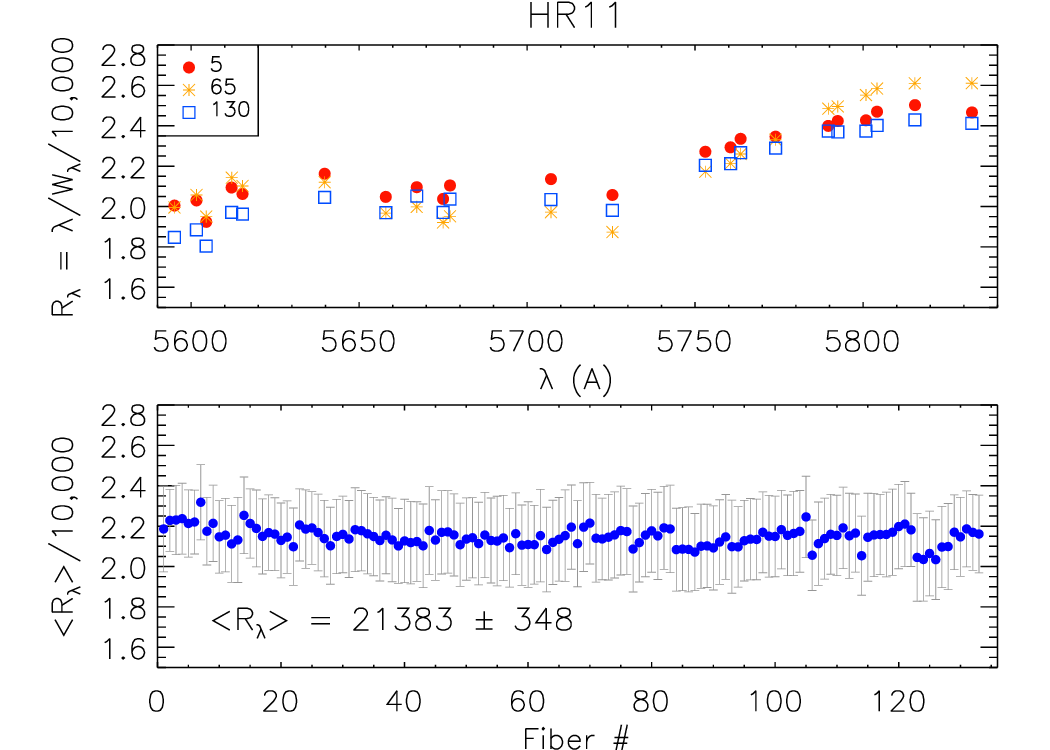}  
\hspace{-1cm}
\includegraphics[width=6.75cm]{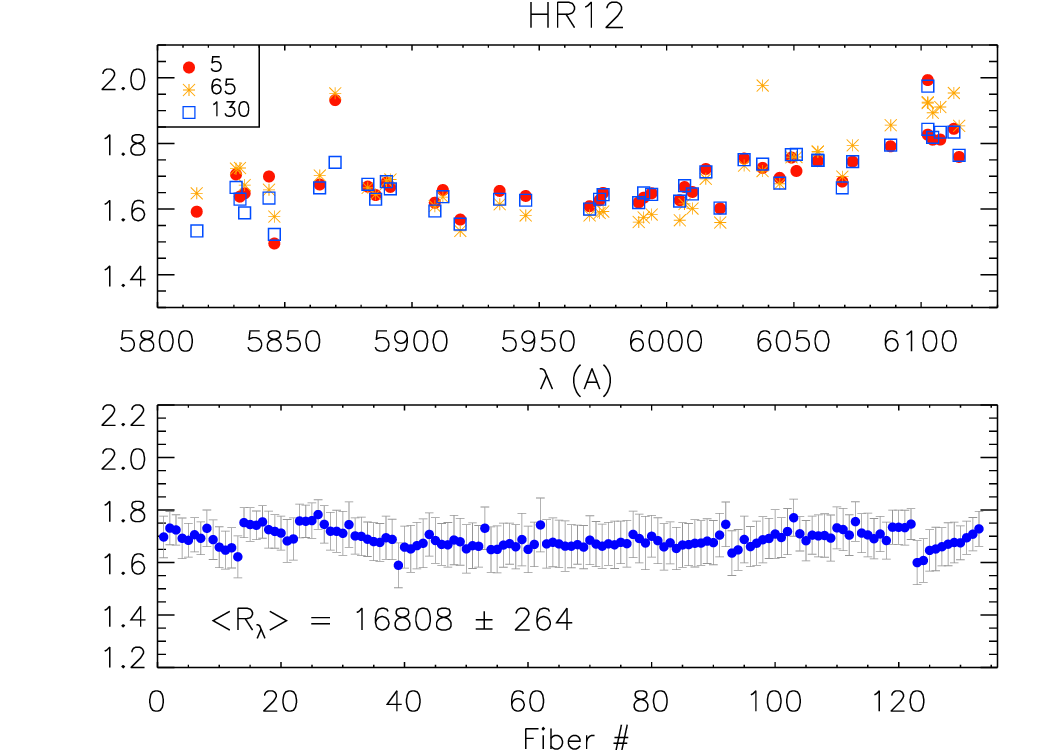}
\hspace{-1cm}
\includegraphics[width=6.75cm]{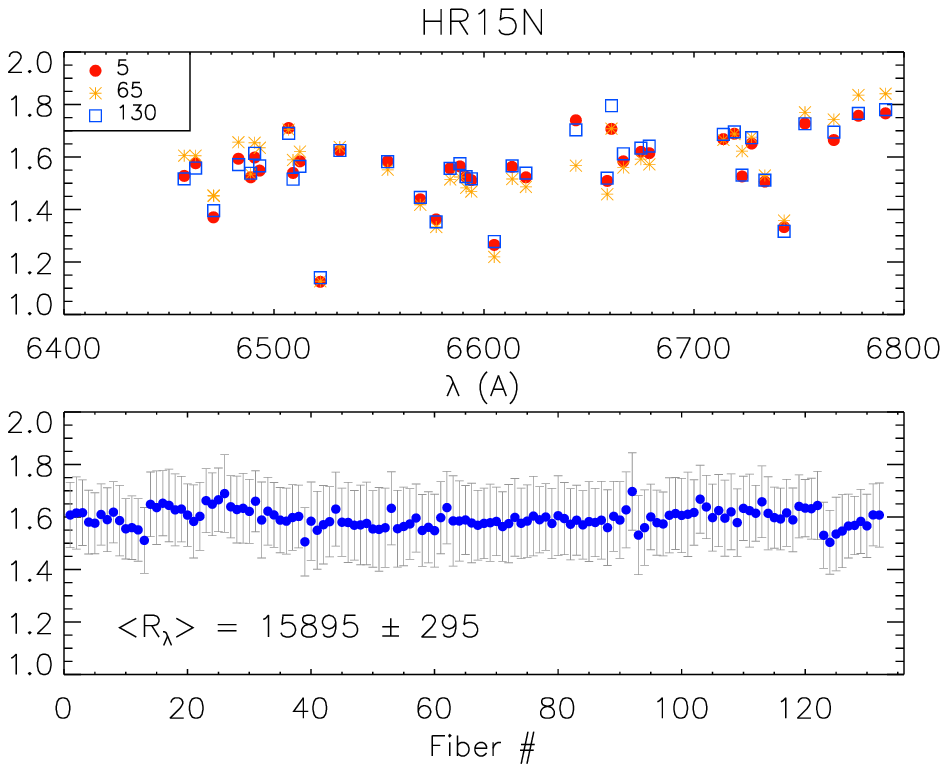}
\caption{Resolving power $R_{\lambda}=\lambda/W_{\lambda}$ of GIRAFFE in the
three setups used for our observations, HR11, HR12, and HR15N, from the left
to the right.  In the upper panels, $R_{\lambda}$, for each setup, is
plotted against wavelength for three fibers near the top, center, and bottom
of the frame.  The lower panels display  the wavelength-averaged value of
$R_{\lambda}$ as a function of the fiber number.  The mean $R$ and its
uncertainty is also reported in each of the lower boxes. }
\label{Fig:Resolution}
\end{center}
\end{figure*}

To obtain accurate measurements of the rotation velocities with \rotfit\ removing
systematic instrumental effects, it is necessary to know the true resolution
of the acquired spectra, which can vary with the wavelength, between one
fiber and another, and in time.  This could be, for instance, the result of
instrument de-focusing.  To this aim, we used the spectra of
wavelength-calibration arc lamps acquired in the same night as the science
frames, selecting unblended emission lines with enough signal that are well
distributed in wavelength (22, 45, and 45 lines for HR11, HR12, and HR15N,
respectively).  We fitted them with Gaussians, which reproduce the
instrumental profile in that position of the focal plane.  The resolution
power is defined, for each line, as $R_{\lambda}=\lambda/W_{\lambda}$, where
$W_{\lambda}$ is the full width at half maximum (FWHM) of the line at wavelength
$\lambda$.  We report these values in Fig.\,\ref{Fig:Resolution} for the
three setups as a function of $\lambda$, in the upper panels, and versus the
fiber number, in the lower panels.  We note a smooth and slight variation of
the resolution with a wavelength of less than $\pm 10$\%, which is apparent in
the HR11 spectra.  However, because of the low nonuniformity of the
resolving power and the difficulty of dealing with spectra with 
wavelength-dependent resolution, we took the mean value and its standard
deviation for each fiber.  This value does not change appreciably with the
fiber number (as shown in the lower boxes of Fig.\,\ref{Fig:Resolution}) and
therefore we took the weighted mean (inverse variance weight) of the
resolution on all fibers and adopted the standard error of the weighted mean
as its uncertainty.  We find the following resolution values, $R_{\rm
HR11}=21,380\pm350$, $R_{\rm HR12}=16,810\pm260$, and $R_{\rm
HR15N}=15,960\pm280$.  The ELODIE templates have been downgraded from their
original resolution of $R_{\rm ELODIE}=42,000$ to match that of the GIRAFFE
spectra when running the code ROTFIT for the simultaneous determination of
atmospheric parameters, radial velocity, and \vsini\
(Sect.\,\ref{sec_rotfit})\footnote{To this aim, we have convolved the
ELODIE templates with a Gaussian kernel of FWHM
$W=c\sqrt{1/R_{\rm GIRAFFE}^2 - 1/R_{\rm ELODIE}^2}$\,\kms.}.

\begin{figure}[htb] 
\includegraphics[width=9cm]{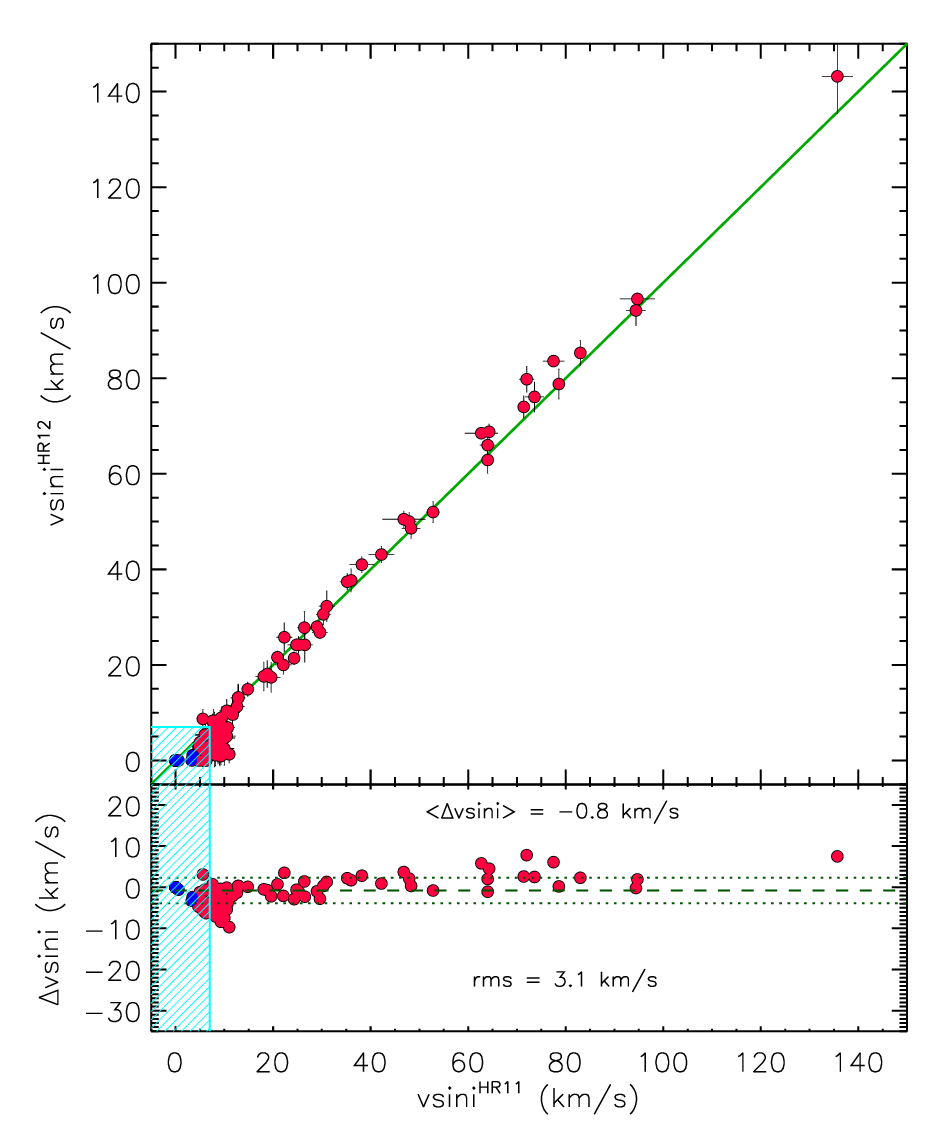}
\caption{Comparison between \vsini\ measured with \rotfit\ on the HR11 and
HR12 GIRAFFE spectra. Red and blue dots are used for main sequence and
giant stars, respectively.  The one-to-one relation is shown by the full
green line in the upper panel.  The differences are displayed in the bottom
panels along with the average value  (dashed lines) and the standard
deviations (dot-dashed green lines).  The cyan hatched areas mark the upper
limit of 7\,\kms\ in both axes.}
\label{Fig:vsini} \end{figure}

For the UVES spectra we checked the spectral resolution in some spectral
range confirming the nominal value of $R_{UVES}=47,000$.  In this case,
because the UVES resolution was slightly larger than  that of the ELODIE
templates, we brought the UVES spectra to the resolution of the ELODIE
ones, by convolution with a Gaussian kernel of $W\approx 3$\,\kms,
although this has a very small effect on the target spectra.
The \vsini\ derived by \rotfit\ for each spectral segment analyzed is
the weighted average of the values that minimize $\chi^2$ for the best
ten templates, where the weight is $w_i=1/\chi_i^2$.  For UVES (and for the
stars observed with more than one GIRAFFE setup), the results from the
individual spectral segments have been averaged, as already mentioned in
Sect.~\ref{sec_rotfit}.  The \vsini\ error has been estimated as the standard
deviation of the individual values to which the uncertainty in resolution
($c\Delta R_\lambda/R_\lambda$) has been added in quadrature.  As a quality
check of the values of \vsini\ derived with \rotfit, we show in
Fig.~\ref{Fig:vsini} the comparison of those derived from spectra with HR11
and HR12 setups.  The average difference between the two datasets is almost
null ($-0.8$\,\kms) and the rms scatter of 3.1\,\kms\ is comparable to the
\vsini\ errors.  

Moreover, as demonstrated by \citet{Frasca2015} through Monte Carlo simulations, the spectral resolution and sampling of the GIRAFFE spectra preclude the measurement of \vsini\ values below 7\,\kms. Consequently, any value yielded by the code below this threshold must be regarded as an upper limit (i.e., \vsini\,$<$\,7\,\kms), as indicated by the cyan hatched area in Fig.~\ref{Fig:vsini}. The upper limit for UVES spectra with the U580
setup has been evaluated as 3\,\kms\ by \citet{Frasca2015}.

\section{Results}\label{sec_APs}

As mentioned above (Sect.~\ref{sec_analysis}), in this work we are approaching the study of NGC\,2509 using a dual methodology, which will be used in the analysis of the WEAVE data. In this section we discuss the results obtained with both procedures in order to test their consistency. Subsequently, we compare our results with the literature, when available.

\subsection{Stellar parameters: Comparison between SP\_Ace and ROTFIT}
The comparison of the APs derived with \Space\ and \rotfit\ is shown in
Fig.\,\ref{Fig:SPACE_ROTFIT}. The \teff\ values obtained with the two
pipelines are in good agreement with each other, with most discrepancies regarding the hottest and faster rotating
stars, as expected.  
The \logg\ values for the giants derived by \Space\ are slightly larger ($\approx$\,0.5\,dex) than the \rotfit\ ones.  
For a few stars classified by \rotfit\ as MS
(\logg$\geq4.0$), \Space\ found low surface gravities.  However, these are
moderately to fast rotating stars, which can explain the discrepancy.  
By considering only single stars with spectra S/N$>$40 and \vsini\ $<10$\,\kmsec\ 
(high-quality measures) the residuals \Space\ - \rotfit\  are $\Delta$\teff\ = $79\pm115$~K, $\Delta$\logg\ = $0.34\pm0.25$, and 
$\Delta$[Fe/H] = $0.04\pm0.07$~dex. On the other hand, when the constraints in S/N and \vsini\ are relaxed, we have $\Delta$\teff\ = $-19\pm257$~K, $\Delta$\logg\ = $-0.03\pm0.33$, and 
$\Delta$[Fe/H] = $0.09\pm0.19$~dex (see Fig.~\ref{Fig:SPACE_ROTFIT}). The larger dispersions of the observed residuals are due to the \Space\ measurements of the stars with large \vsini. These are all dwarf stars observed with GIRAFFE, most of them having spectra with low S/N. In addition, the wide rotational line profile of such spectra cannot be correctly fitted by \Space, which tends to deliver unreliable results in such cases. When high-quality spectra are taken into account, these show small systematics in \teff, \logg, and [Fe/H]. Since \rotfit\ and \Space\ are based on different references (EWs synthesis based on Kurucz atmosphere models for \Space, real observed spectra and stellar parameters from the ELODIE spectral library for \rotfit), different zero points between the two are expected. In fact, we note that methods based on atmosphere models and synthesis, such as \Space\ and SYNTHE, are in good agreement (up to 0.01~dex) for [Fe/H], which is significantly better than the systematic offset of 0.04~dex between \Space\ and \rotfit\ for high-quality measures.

In Table~\ref{t:TGM_ABD_SPACE} we report the stellar parameters and chemical abundances derived by \Space.
We observe a slight underestimation of  \logg\ in dwarf stars, as expected from the
quality tests reported in \citet{Boeche2021}, but also an overestimation of
\logg\ for giant stars.
This comparison emphasizes once again the role played by the analysis method
and the necessity of understanding where each method performs better and
what are the biases and offsets.  In particular, both \Space\ and ROTFIT
will be used for FGK OC stars observed by the WEAVE surveys \citep{Jin2024}. In this framework, it is important to test the consistency of the results of the two pipelines.

\subsection{Chemical abundances: \Space\ and SYNTHE}

The \Space\ chemical abundances\footnote{\Space\ abundances are based on \citet{grevessesauval1998} solar abundances. However, to be consistent with SYNTHE measurements, we converted them to the \citet{Grevesse07} solar abundances.} [X/H] are consistent among the OC stars.
As expected, the UVES spectra provide higher precision
relative to GIRAFFE ones, with standard deviations of chemical abundances of the cluster smaller than 0.1~dex for all elements in the UVES spectra with \logg\ $< 3.5$ (see Table~\ref{tab_abb_aver}). The mean [X/Fe] of five
s-process elements (Y, Zr, Ba, La, Ce) significantly deviates from the solar one, as clearly visible in Fig.~\ref{f:ABD_SPACE}. This also applies to the elements Na and Al.
It is also worth noting that the abundances obtained from dwarf stars show an excellent agreement with those obtained from giant stars. The only exception is La, whose difference reaches $\approx$\,0.2\,dex.
Regarding the SYNTHE abundances, NGC\,2509 exhibits a solar composition for most of the elements. Only for some of them (Na, Zn, and Ce) the values are slightly supersolar ([X/H]\,>\,0.2\,dex).

The average composition of the cluster, [X/H], is also reported in Table~\ref{tab_abb_aver}. 
For each element, it has been calculated by taking the average of
the values for each star, and the uncertainties express (in terms of standard deviation) the dispersion of
stellar abundances around the cluster value.
Both sets of abundances, SYNTHE and \Space, agree within 0.1\,dex. In just some cases (Al, Si, Ti, and La) the differences are higher, up to 0.15\,dex. Heavy elements such as Ce and Ba show the greatest contrasts (0.23 and 0.28\,dex, respectively).
The dispersion of [Fe/H], [X/H], and  [X/Fe] is slightly higher than in the case of \Space\ determination. 
\begin{table}[ht]
\caption{Average chemical abundances for NGC\,2509, relative to solar abundances 
by \citet{Grevesse07}, obtained from giant stars observed with SYNTHE and \Space. }
\label{tab_abb_aver}
\begin{center}
\begin{tabular}{lr|rr}  
\hline\hline
X    &  SYNTHE~ &  \Space\ ~~   \\ %&  [X/Fe]\\
     &  [X/H]~(dex)   & [X/H]~(dex) \\
\hline
Na   &     0.30 $\pm$ 0.04  &    0.40 $\pm$ 0.04  \\ %& 0.25\\
Mg   &  $-$0.04 $\pm$ 0.07  &    0.03 $\pm$ 0.03  \\ %& $-$0.09\\
Al   &     0.12 $\pm$ 0.04  &    0.25 $\pm$ 0.03  \\ %& 0.14\\
Si   &     0.05 $\pm$ 0.07  &    0.17 $\pm$ 0.03  \\ %& 0.04\\
Ca   &     0.09 $\pm$ 0.05  &    0.04 $\pm$ 0.02  \\ %& $-$0.08\\
Sc   &     0.08 $\pm$ 0.07  &    0.06 $\pm$ 0.02  \\ %& $-$0.06\\
Ti   &     0.08 $\pm$ 0.07  &    0.19 $\pm$ 0.03  \\ %& 0.07\\
V    &     0.11 $\pm$ 0.05  &    0.18 $\pm$ 0.02  \\ %& 0.06\\
Cr   &     0.12 $\pm$ 0.06  &    0.08 $\pm$ 0.03  \\ %& $-$0.05\\
Mn   &     0.13 $\pm$ 0.05  &    0.03 $\pm$ 0.02  \\ %& $-$0.09\\
Fe   &     0.11 $\pm$ 0.06  &    0.12 $\pm$ 0.03  \\ %&   \\
Co   &     0.21 $\pm$ 0.06  &    0.15 $\pm$ 0.03  \\ %& 0.02\\
Ni   &     0.13 $\pm$ 0.06  &    0.13 $\pm$ 0.03  \\ %& 0.01\\
Cu   &                      &    0.06 $\pm$ 0.03  \\ %& $-$0.06\\
Zn   &     0.28 $\pm$ 0.04  &                     \\ %     & \\
Y    &     0.13 $\pm$ 0.06  &    0.22 $\pm$ 0.02  \\ %&  0.10\\
Zr   &     0.15 $\pm$ 0.04  &    0.22 $\pm$ 0.04  \\ %& 0.10\\
Ba   &     0.15 $\pm$ 0.07  & $-$0.12 $\pm$ 0.03  \\ %& $-$0.24\\
La   &     0.18 $\pm$ 0.08  &    0.32 $\pm$ 0.03  \\ %& 0.19\\
Ce   &     0.22 $\pm$ 0.09  &    0.00 $\pm$ 0.06  \\ %& $-$0.12\\
Nd   &     0.19 $\pm$ 0.07  &    0.17 $\pm$ 0.03  \\ %& 0.05\\
\hline
\end{tabular}
\end{center}
\end{table}

\subsection{Comparison with the literature}
To our knowledge, only one RC star ($Gaia$-DR3  5714209934411718784) has published APs and abundances based on high-resolution spectroscopy. This star is part of the OC sample of the SPA project (i.e., the Stellar Population Astrophysics Large Program, PI L. Origlia, done at the Italian Telescopio Nazionale Galileo). \citet{Zhang21} derived T$_{\rm eff}$=4705~K, \logg=2.53, and [Fe/H]=$-0.10$\,dex. These values were recently updated by \citet{marina25}  to T$_{\rm eff}$=4773$\pm$39~K, \logg=2.885$\pm$0.104, and [Fe/H]=0.212$\pm0.014$\,dex.
It must be noted that \citet{marina25} employed a full non-LTE analysis unlike \citet{Zhang21} and the present work in which LTE is assumed.
The star is not among the FLAMES targets; however, if we take the ten RC stars observed with UVES, the average ROTFIT values for T$_{\rm eff}$, \logg, and [Fe/H] are in agreement (4773\,K, 2.75, and 0.062\,dex, respectively, with standard deviations of 89\,K, 0.09, and 0.04\,dex, respectively).

In addition, NGC~2509 stars were observed with the $Gaia$ Radial Velocity Spectrometer (RVS; see, e.g., \citealt{recio2023}) and
14 of our targets have $Gaia$ RVS stellar parameters. The means and standard deviations for the iron abundance and alpha elements of the $Gaia$ RVS data are ${\rm [M/H]} = 0.16\pm0.19$~dex and $[\alpha/{\rm Fe}] = -0.14\pm0.09$~dex. In comparison, the  \Space\ results provide ${\rm [M/H] }= 0.09\pm0.03$~dex and $[\alpha/{\rm Fe}] = -0.038\pm0.004$~dex (where the metallicity was computed following the formula of \citet{salaris1993} and $[\alpha/{\rm Fe}]$ is the mean of the abundances of [Mg/Fe], [Si/Fe], and [Ca/Fe]). The metallicities are in good agreement while the alpha abundances stand apart, although of just one sigma. Regarding the ROTFIT plus SYNTHE results, $[\alpha/{\rm Fe}] = -0.07\pm0.07$~dex if we use the same elements (Mg, Si, Ca), so the resulting [M/H] is 0.06$\pm0.06$. Again, the agreement is within one sigma. %}
From the $Gaia$ RVS measurements we also computed the mean radial velocity of the cluster from 22 stars obtained after pruning the sample from the SB2 stars and the one with $RV_{err} > 2$~\kmsec. With this clean sample, the mean radial velocity of the cluster is $RV_{NGC2509} = 60.8\pm1.2$~\kmsec.
This value is in good agreement with the result reported in Sect.~\ref{sec_RV}. \\

\section{The age of NGC\,2509}\label{sec_age}
NGC\,2509 is a moderately old OC whose age, even in recent studies, remains somewhat uncertain and ranges from about 900\,Myr \citep[][who used stellar models with rotation, which implies lower ages]{deJuanOvelar2020} to about 1.2\,Gyr \citep[][who employed isochrones without rotation]{carraro2007} to about 1.6\,Gyr according to large OC surveys such as \citet{cantat2020} and \citet{HR2024}, who used artificial and convolutional neural networks, respectively.
In order to shed some light on this question, we investigate the age of NGC\,2509 using two different approaches, based on the work done in the current research.

\subsection{Content of Lithium}
\begin{figure}[]
\hspace{-.3cm}
\includegraphics[width=9.8cm]{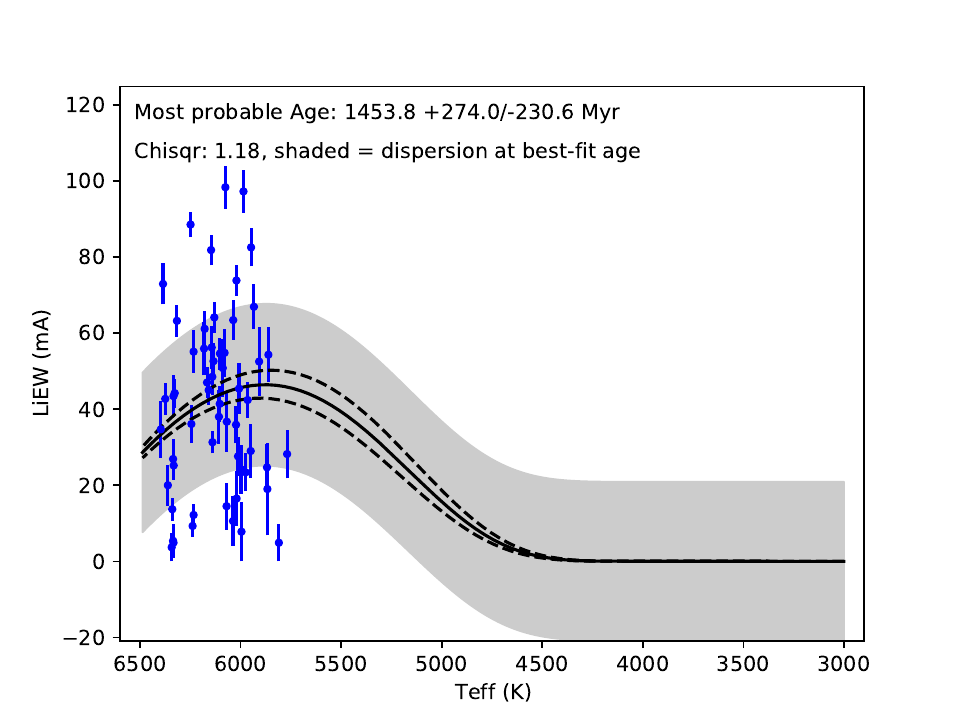}
\caption{Determination of the cluster age from the Li-depletion pattern.}
\label{eagles}
\end{figure}

The first method is the evaluation of the abundance of Li in MS stars, since it is known to be age-dependent \citep[see, e.g.,][]{Soderblom83,Navascues04,Pleiadi}. Studying its distribution among the cluster members as a whole provides a good estimate of the cluster age. We followed the same procedure used in previous works; for a detailed description, the reader is referred to \citet{Pleiadi}, in which we applied the same method to the Pleiades. In a first step, we measured the EW of the \ion{Li}{i}\,6708\,\AA\ line by spectral subtraction. Then, using the code EAGLES \citep{eagles}, the age of the cluster is automatically determined after finding the model that best reproduces the Li-depletion pattern displayed in the \teff-EW(Li) diagram. In our analysis, we used the MS stars in the temperature range in which the code operates (3500$\leq$\teff$\leq$6500\,K), counting a total of 57. Their EWs are reported in Table~\ref{tab_eqw}. However, our sample is mainly distributed in the 6000--6500\,K interval, since the coolest stars observed are giants. The resulting age, whose fit is shown in Fig.~\ref{eagles}, is $\approx$1450$\pm$250\,Myr.  

\subsection{Isochrone fitting and reddening}\label{sec_isoc}
In this section we derive the age by isochrone fitting.
The CMD is fitted using the Bayesian approach described in \citet{2025A&A...Kallimanis}. 
To this aim, we first built the CMD based on the $Gaia$-DR3 G, BP, RP photometry. 
In addition, we make use of the precise and well-verified synthetic photometry derived from $Gaia$ XP spectra by \citet{montegriffo23} in the $g,i,z$ SDSS bands. This  multi-band approach has proven to alleviate the degeneracy between metallicity, extinction, age, and distance because  different passbands have different sensitivity to these parameters (see, among others,  \citealt{2022A&A...658A..91A, 2023A&A...669A.104K, 2024AJ....167...12C}).
We assume as prior our determination of [M/H]=0.12\,dex.
Regarding the distance modulus, we utilize a prior derived from the
$Gaia$-DR3 parallaxes of the cluster stars.
No correction is applied for the zero-point offset described in \citet{Lindegren21}.
The extinction is derived in terms of the extinction parameter A$_0$ of the Fitzpatrick extinction law \citep{fitzpatrick1999} and then converted in terms of $E(BP-RP)$ and $A_G$.  In the case of large passbands such as $G$, and $BP,RP$, the extinction coefficients  depend on the intrinsic temperature (or color) of the star. We make use of the empirical coefficients derived by \citet{2018A&A...614A..19D}, where a discussion of the problem is presented.
We make use of the PARSEC
isochrones with rotation \citep{2022A&A...665A.126N}. Stellar rotation is known to affect both the
luminosity and the color of a star. In addition, rotating stars have rotationally induced mixing, which increases  luminosities and  lifetimes. Centrifugal forces  introduce a latitudinal dependence on the effective surface gravity and on the effective surface temperature of the star. This implies that the observed color of a rotating star changes with the stellar axis inclination angle. Here we assume  that the instantaneous (not initial) rotational velocity compared to the break-up one, $\Omega/\Omega_{crit}$, ranges from 0.1 to 0.9, although  our \vsini\ determinations suggest values larger than 0.5-0.6 (see Sect.~\ref{sec_emsto} for a discussion). We used the magnitudes  derived from the flux averaged over all inclinations assuming an isotropic distribution of inclinations.

\begin{figure*}[]
\centering
\includegraphics[width=16cm]{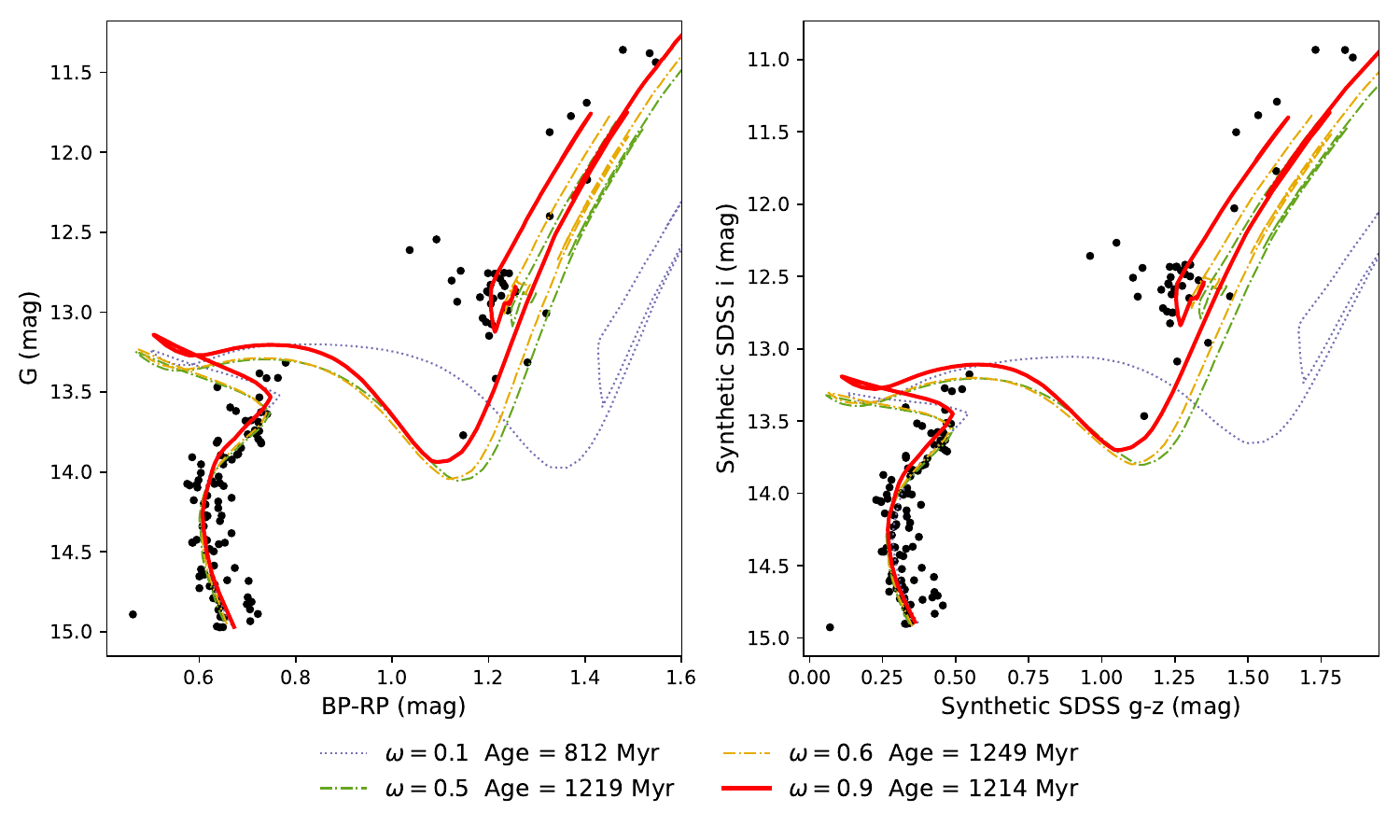}
\caption{$Gaia$ CMDs of NGC\,2509 with the best-fitting isochrones at different ratios $\Omega/\Omega_{crit}$ in the $G$, ($G_{\rm{BP}}-G_{\rm{RP}}$) passbands (left) and SDSS ($i, g-z)$ synthetic photometry (right). Best fit was obtained for $\Omega/\Omega_{crit} = 0.9$, plotted in red.}
\label{CMD_isoc}
\end{figure*}

Then we perform a Bayesian fit for the CMDs in $(G, BP-RP)$ and synthetic
SDSS $(i, g-z)$ simultaneously using the ASteCA package \citep{Perren_2015}
as implemented by \citet{2025A&A...Kallimanis}, optimizing
the logAge and distance modulus ($\mu$) parameters. 
The CMD cluster clearly shows the sequence of equal mass binaries.
ASteCA deals with the presence of binaries by assuming   the \citet{dkmassratios}  distribution 
of binary mass ratios across the isochrone. 
The percentage of binary stars is assumed to be uniform within the cluster
and is also sampled (flat prior between 0 and 1). However, this
does not correspond to the true binary fraction of the cluster.
Indeed, the real distribution of binary fractions across masses is not exactly uniform as assumed here, but depends on the mass of the stars \citep{offner}. Nevertheless, it is included in the fit as a nuisance
parameter.
Figure~\ref{CMD_isoc} presents the CMD fit at varying $\Omega/\Omega_{crit}$, while Fig~\ref{CMD_solution} shows the quality of the fits. In Table \ref{tab:fit_results} the results of the isochrone fits are summarized.

Values of $\Omega/\Omega_{crit}$ smaller than 0.6 fit the magnitude and color of the main sequence, but do not reproduce the color of the red clump, and the solution is bimodal in age.
When $\Omega/\Omega_{crit}=0.9$ we obtain the highest Bayesian evidence. In this case, we derive an age of 1214$^{+57}_{-30}$\,Myr, $\mu=12.13 \pm 0.04$\,mag, and  $A_0=0.20 \pm 0.03$\,mag.
 When $\Omega/\Omega_{crit}=0.1$, we obtain the youngest age, 812$^{+371}_{-45}$  Myr. 
The oldest age is derived for $\Omega/\Omega_{crit}=0.6$  and is 1249  Myr. 
All these age values are compatible within the errors.  The quoted
uncertainties are the internal errors of the fit and provide a lower limit
to the true uncertainties.  The fact that $\Omega/\Omega_{crit}$ is not
known

introduces an overall age uncertainty of $437$\,Myr.
 The distance modulus goes from 12.23$^{+0.07}_{-0.17}$ to 12.04 $^{+0.04}_{-0.03}$ and the extinction goes from $A_0=0.20 \pm 0.03 $ to $0.79^{+0.05}_{-0.40}$. 

We note excellent agreement between the results obtained from these two approaches (abundance of Li and isochrone fitting). This result is compatible with \citet{cantat2020} and \citet{HR2024}, while it is slightly older than the value of about 900 Myr derived by  \citet{deJuanOvelar2020}. While we note that the isochrone fit we obtain reproduces all the features of CMD well, we point out that their fit is obtained using MIST isochrones \citep{Dotter2016}.  It is well known that the behavior of stellar  isochrones including rotation is dependent on the model assumptions, in particular concerning  the convective and rotational mixing efficiencies \citet{martinelli2025}.  As described in \citet{2025A&A...701A.258N}, the PARSEC 2.0 overshooting implementation uses ballistic step, while MIST adopts 
exponential decay. Convective regions are defined using Schwarzschild (PARSEC 2.0) versus
Ledoux (MIST).  In PARSEC v2.0, the diffusion
scheme is  used to treat meridional circulation  and shear instability, while in MIST
models  five different rotationally induced instabilities  are taken into account. These differences result in different impacts of rotational mixing. This could explain the age difference, since rotating PARSEC tracks tend to be  more luminous at the turnoff.

\begin{figure}[]
\includegraphics[width=\columnwidth]{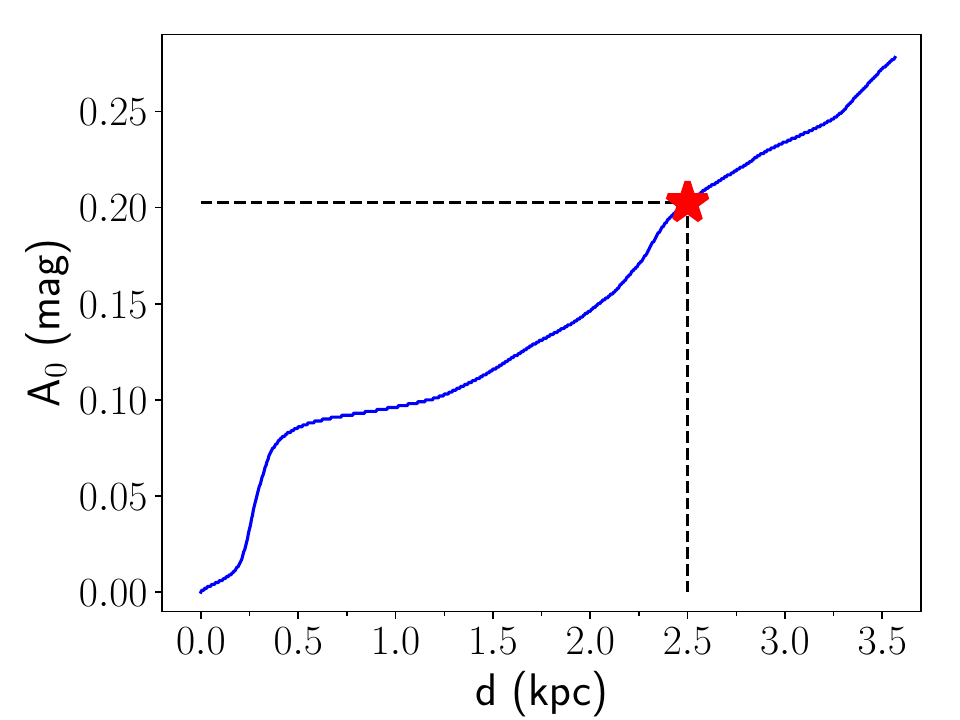}
\caption{Extinction ($A_0$) in the line of sight of NGC\,2509 (blue line) from \citet{Lallement19}. The expected value for the cluster, according to its position, is marked with the star.}
\label{fig_reddening}
\end{figure}

We compare our results with  the extinction parameter $A_0$ from the $Gaia$-DR3 data. 
The parameter $A_0$ is provided for nearly all stars in our sample observed by $Gaia$, however, these values are not sufficiently precise to be used to correct for individual reddening, but can be averaged over the region to obtain a mean value \citep{2023A&A...674A..32B}. We find an average value of $A_0$=0.15 $\pm 0.14$\,mag on 128 stars. 
We also resorted to the Galactic 3D dust maps\footnote{Interactive version available at: \url{https://astro.acri-st.fr/gaia_dev/}} implemented by \citet{Lallement19} based on 2MASS and $Gaia$-DR2 data. As displayed in Fig.~\ref{fig_reddening}, the extinction expected at the cluster distance is about $A_0$= 0.2\,mag.
Both results are in agreement with our highest Bayesian evidence determination. 

\begin{table}[t]
\centering
\caption{Isochrone fit results for NGC~2509. }
\label{tab:fit_results}
\begin{tabular}{c || c c c | c}
\hline
\hline
$\Omega/\Omega_{crit}$ & Age (Myr) & $A_V$ & $\mu$ & $\log Z$ \\
\hline
&&&&\\
0.1 & $812^{+371}_{-45}$ & $0.79^{+0.05}_{-0.40}$ & $12.23^{+0.07}_{-0.17}$ & $573.9$ \\
&&&&\\
0.5 & $1219^{+35}_{-21}$ & $0.31^{+0.03}_{-0.03}$ & $12.05^{+0.05}_{-0.04}$ & $578.8$ \\
&&&&\\
0.6 & $1249^{+17}_{-24}$ & $0.27^{+0.02}_{-0.02}$ & $12.04^{+0.04}_{-0.03}$ & $584.9$ \\
&&&&\\
0.9 & $1214^{+57}_{-30}$ & $0.20^{+0.03}_{-0.03}$ & $12.13^{+0.04}_{-0.04}$ & $602.0$ \\
&&&&\\
\hline
\end{tabular}
\tablefoot{The last
column is the Bayesian evidence, which increases for higher
values of $\Omega/\Omega_{crit}$, indicating a better fit quality.}

\end{table}

\section{Discussion} \label{discussion}

\subsection{Galactic metallicity gradient}

\begin{figure}[ht]
\begin{center}
\includegraphics[width=\columnwidth]{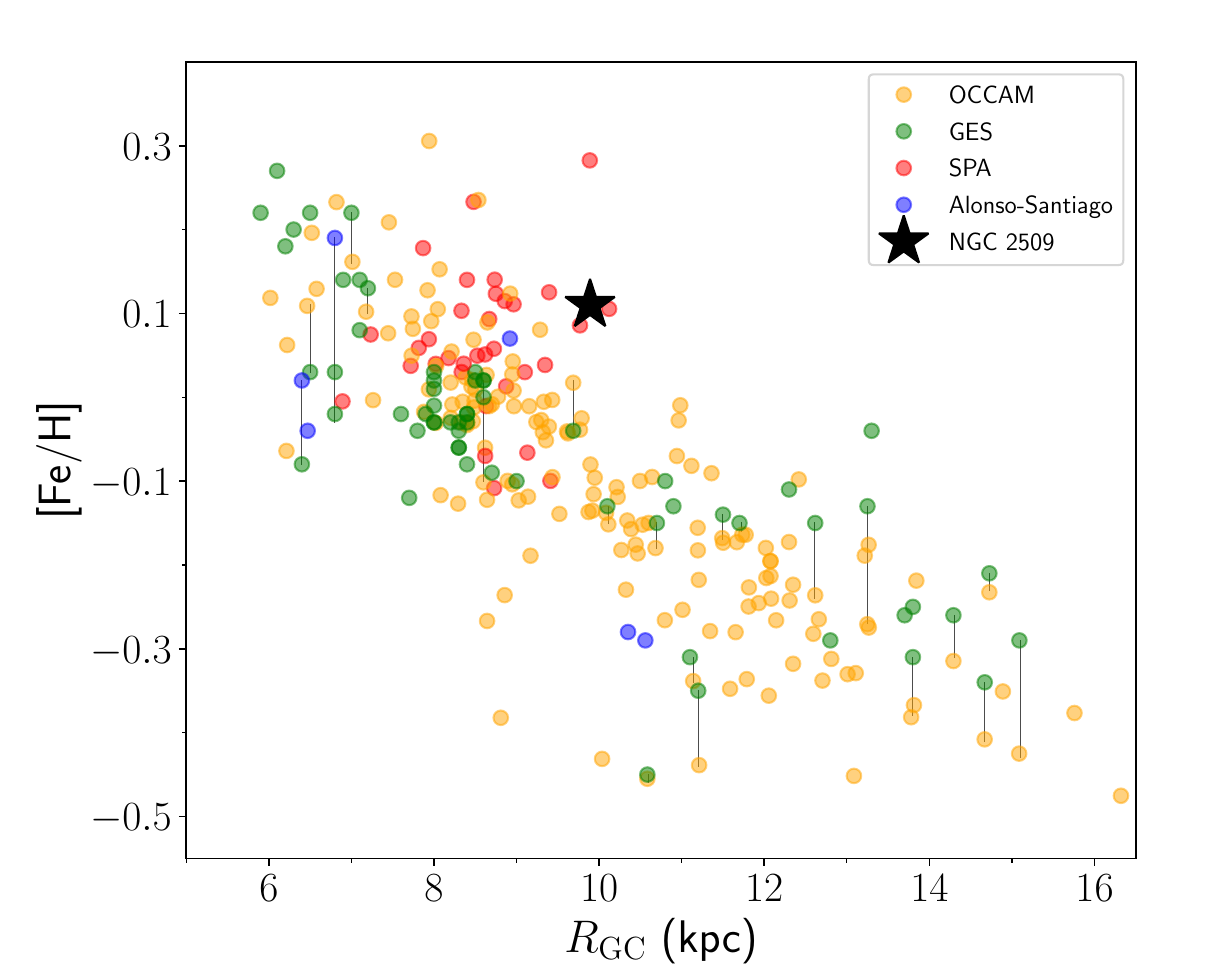}   
\caption{Radial metallicity gradient traced by open clusters studied with high-resolution spectroscopy. Black lines link results for the same cluster provided by different authors. The star represents the metallicity found in this work for NGC\,2509.}
\label{fig_grad}
\end{center}
\end{figure}

Open clusters are recognized as excellent indicators for mapping the radial distribution of metallicity within the Galaxy, often referred to as the Galactic gradient. To assess how the metallicity determined for NGC\,2509 in this study aligns with this gradient, we compiled a sample of clusters from the literature that had been analyzed using consistent methods. Our compilation includes metallicities derived from high-resolution spectroscopy as part of the $Gaia$-ESO \citep{randich2022} and OCCAM \citep[APOGEE-DR19;][]{otto26} surveys. 
We further enriched this sample with clusters investigated by or related to the SPA project
\citep{Frasca2019,Stock2,M39,Radcliffe,marina25} and several other young open clusters from our own research group \citep{6067,3105,2345,3OC}. In total, our dataset comprises almost 200 clusters, spanning galactocentric distances ($R_{\textrm{GC}}$) up to 16\,kpc. Figure \ref{fig_grad} illustrates the position of NGC\,2509 relative to this gradient. The iron abundance, representing metallicity, was normalized to A(Fe)=7.45 \citep{Grevesse07}. Galactocentric distances were adopted from \citet{cantat2020}, who calculated them using $Gaia$-DR2 astrometry and assuming a solar reference distance of $R_{\odot}$=8.34\,kpc.
NGC\,2509 lies in the upper envelope of the gradient, which means that its metallicity is somewhat higher than the mean value observed in other clusters at a similar distance.

\subsection{Chemical composition and Galactic trends}

\begin{figure*}[ht]
\centering
\includegraphics[width=17cm]{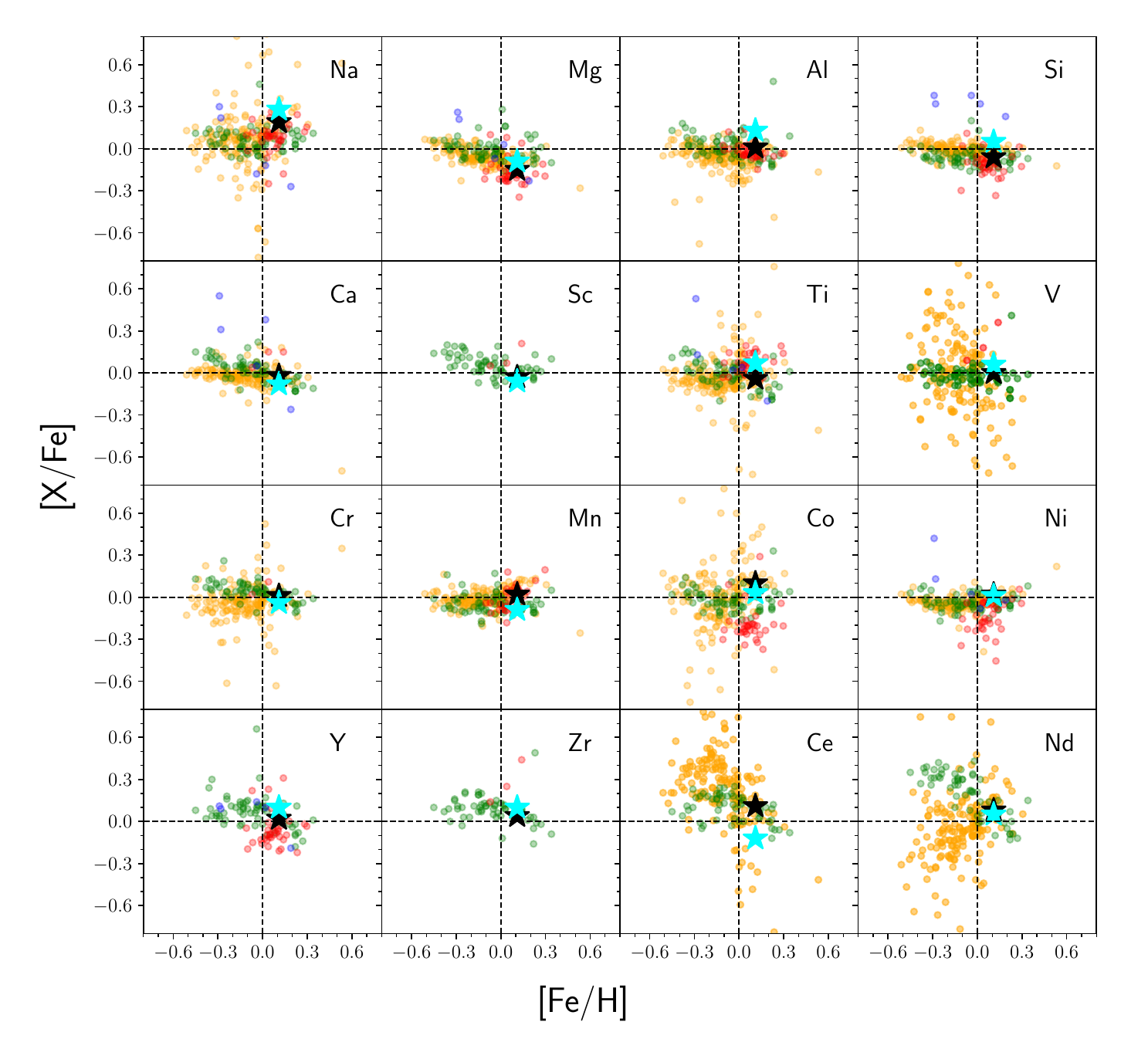}
\caption{Abundance ratios [X/Fe] vs. [Fe/H] for NGC\,2509. The black stars represent the values derived with SYNTHE while the cyan ones indicate those calculated with \Space. The remaining symbols and colors are the same as in Fig.~\ref{fig_grad}. The dashed lines show the solar value. 
}
\label{fig_trends}
\end{figure*}

As noted in Sect.\ref{sec_intro}, this work marks the first determination of chemical abundances for NGC\,2509 of a large sample of stars. Lacking direct comparative literature, we assessed our findings against the chemical trends observed in Galactic OCs. The cluster sample used for the Galactic gradient discussion was re-employed for this purpose. Notably, for this analysis, chemical abundances for GES clusters were drawn from \citet{Magrini2023}. We compared the chemical composition of NGC\,2509 with that of the other OCs in our collected sample. Figure \ref{fig_trends} displays the [X/Fe] versus [Fe/H] ratios for the 16 chemical elements in common \citep[scaled to solar abundances from][]{Grevesse07}. The abundances of NGC\,2509 show excellent agreement with the established trends of the other OCs, leading us to conclude that its chemical composition is fully compatible with the Galactic thin disk. 
  
\subsection{On the lack of an eMSTO in NGC~2509 }\label{sec_emsto}

\begin{figure}[]
\centering
\includegraphics[width=9cm]{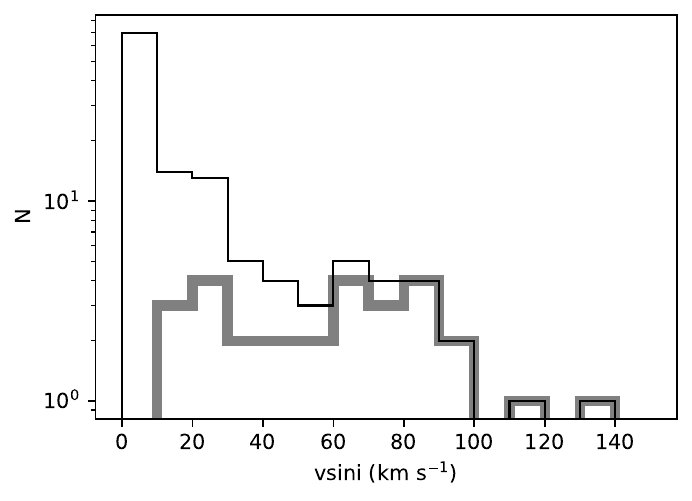}
\caption{Histograms of the \vsini\ for the whole sample (thin black line) and for stars belonging to the TO only (thick~gray line).}
\label{f:vsini_MSTO}
\end{figure}

\begin{figure}[]
\centering
\includegraphics[width=9cm]{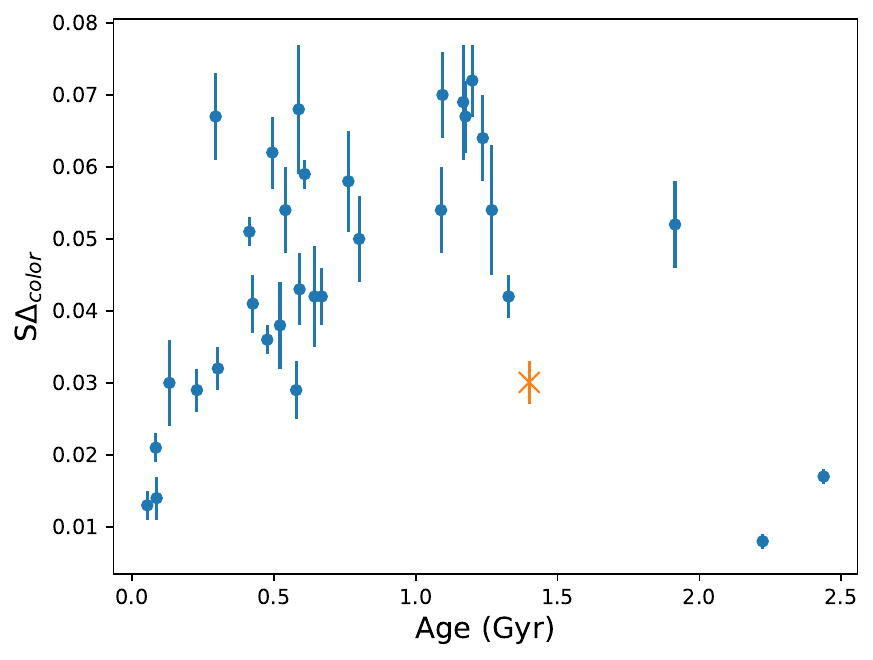}
\caption{Age (x-axis) versus S$\Delta_{color}$ (y-axis) of the MSTO for NGC~2509 (orange cross) and values by \citet{cordoni24} of the other 32 OCs MSTO (blue points).}
\label{f:deltacolor}
\end{figure}

As discussed in Sect.~\ref{sec_intro}, the eMSTO phenomenon has generally been explained as an effect of the stellar rotation \citep{milone,sun, martinelli2025} or as an age spread. The discussion is still open, since none of the proposed mechanisms are fully satisfactory. The age spread scenario requires an age difference on the order of hundreds of Myr \citep{2025A&A...701A.221S}. This is in contrast to the current understanding of cluster formation. A rapid dispersal of the remaining gas from the stellar feedback will conclude the star formation in less than 10 Myr \citep{1997ApJ...480..235E}.  In addition, the age spread scenario cannot explain some of the CMD features that are compatible with a single population, such as a compact red clump star distribution or the sub-giant branch morphology \citep{2015MNRAS.448.1863B}.  Recent observations have shown that 
both fast and slow rotators are present in the case of eMSTO.  The higher the \vsini, the redder the star appears to be \citep{2018AJ....156..116M}. However, while \citet{2011MNRAS.412L.103G} find that rotation alone cannot explain the eMSTO features, several studies reach the opposite conclusion, i.e., that a convenient distribution of rotation rates  can  qualitatively account for the observed morphology, at least in some cases \citep{2015MNRAS.448.1863B, cordoni24}.
Figure~\ref{f:cmd_vsini} displays the cluster CMD highlighting the \vsini\ for the stars observed spectroscopically in this work. As expected, the fastest rotators are located in the upper MS and MSTO.
Selecting the TO stars with the constraints $BP-RP <  1.0$~mag, $G < 15$~mag, and excluding the binary stars, we see that their distribution in \vsini\ is rather uniform and  in the range of 20--135~\kms\ in contrast with the whole sample for which most stars have \vsini$<40$\kms\ (see Fig.~\ref{f:vsini_MSTO}). 
The range of \vsini\ in this cluster is rather narrow compared to objects
showing an extended turnoff.  Several studies have derived the rotation
velocity in clusters with eMSTO morphology.  \citet{cordoni24} analyze a
sample of 17 eMSTO clusters and find a very broad distribution of  the Gaia
parameter $v_{broad}$ that includes both  rotation and instrumental
broadening effects.  In these objects, $v_{broad}$ at the turnoff ranges
from a few km~s$^{-1}$   up to 300 km~s$^{-1}$ and is sometimes bimodal (see
their Fig.~3).  Similar results are presented by
\citet{2025arXiv251205458R}.  To compare the NGC~2509 eMSTO morphology with
that of Cordoni et al., we used the parameter $S\Delta_{color}$ used by
\citet{cordoni24}.  This parameter estimates the width of the eMSTO in terms
of color extension normalized to the color width of the MS at fainter
magnitudes (see Appendix~\ref{sec_vert_color} for a detailed description). 
In Fig.~\ref{f:deltacolor} we show the $S\Delta_{color}$ value of NGC~2509
eMSTO in comparison with the values of the other 32 OC eMSTOs by
\citet{cordoni24}.  NGC~2509 eMSTO appears to be
narrower than the MSTOs of other OCs of similar age.

Our data do not provide any constraint on the orientation of the rotation axis of the stars. However, assuming sin\,$i$=1, we can use \vsini\ as a proxy for the minimum equatorial rotational velocity ($v_{tan\,eq}$) in the cluster. We compare these values with the expectations from the PARSEC isochrones.  When $\Omega/\Omega_{crit}$=0.5, the turnoff mass at an age of 1219 Myr is 1.74 M$_0$ (and 1.97 M$_0$ at the MS tip), with $v_{tan\,eq}$=94.9 km~s$^{-1}$. If we define the turnoff region as  G$_{abs}=0.62 - 3.19$\ mag, i.e., G=12.67--15.24 (assuming the distance modulus $\mu$ derived in Sect.~\ref{sec_isoc}), we find that $v_{tan\,eq}$  goes from 60 km~s$^{-1}$ to 115 km~s$^{-1}$. When  $\Omega/\Omega_{crit}$ =0.6, $v_{tan\,eq}$ in the turnoff region varies from 84 km~s$^{-1}$ to 143.8 km~s$^{-1}$. Both values of $\Omega/\Omega_{crit}$ are in broad agreement with  the observed \vsini\ range, while the lower values  do not reproduce the observations. 
Comparing our determinations of \vsini\, we can conclude that the minimum possible value of  $\Omega/\Omega_{crit}$ should be in the range [0.5-0.6].   Our highest Bayesian evidence solution, $\Omega/\Omega_{crit}=0.9$ provides $v_{tan\,eq}$ at the turnoff in the range 130-220 km~s$^{-1}$, higher than observed. To reconcile these values with the observed \vsini\, we need to assume lower values of $sin i$. Our results  partially support the tight rotational velocity distribution suggested by \citet{deJuanOvelar2020}, who find that the CMD is in agreement with values of $\Omega/\Omega_{crit}$ in the range  [0.4, 0.6].
Clearly both the CMD fit and the  higher Bayesian evidence value of  $\Omega/\Omega_{crit}$ depend on the  implementation of the rotation in the PARSEC models. In particular, the red clump color and magnitude are only marginally dependent  on  rotation through the mass loss. Indeed, after leaving the main sequence, the
expansion of the envelope results in a decrease in surface
rotation velocity  and the star evolves as a non-rotating object. Instead, the main sequence becomes cooler at increasing $\Omega/\Omega_{crit}$, reducing the difference in color between the main sequence tip and the red clump. This is due to the proportionality between T$^4_{eff}$ and effective gravity, which is reduced in turn by centrifugal force.  In MIST isochrones, the different treatment of rotation and of mixing results in a slightly cooler main sequence phase at similar values of   $\Omega/\Omega_{crit}$. This can explain the lower values of $\Omega/\Omega_{crit}$ obtained by  \citet{deJuanOvelar2020}.

\section{Summary and conclusions}\label{summary}

Open clusters are very interesting from two perspectives: both for the study of stellar evolution and in Galactic archaeology. In this context we investigated NGC\,2509, a little-studied OC that exhibits a narrow distribution of MS stars and whose chemical composition has been studied extensively for the first time in the present work. We performed moderate-resolution spectroscopy ($\approx$\,20\,000) with GIRAFFE for most of the targets in our sample, mainly MS stars. Additionally, we also took high-resolution spectra ($\approx$\,47\,000) of the giants with UVES. 
In total we observed 132 stars, all of them good candidate members according to \citet{HR2024}. Our sample represents a significant fraction of their entire reported member list, more than 70\% of stars with a high probability  (0.7) of membership.
In our analysis, we used two different methodologies (\rotfit + SYNTHE and \Space), both of which will be used in the analysis of the first observations made with WEAVE, which are about to be carried out.

We provide atmospheric stellar parameters (\teff, \logg, [Fe/H]) as well as projected rotational (\vsini) and radial velocities (RV) for all our targets. 
The results of both codes are compatible within the errors. %\corrado{
The systematic differences between \rotfit\ and \Space\ in stellar parameters are 91~K in \teff, 0.3~dex in logg, and 0.09~dex in [Fe/H]. %}
From all single stars, both dwarfs and giants, we find the average cluster RV=58.6$\pm$1.3\,km\,s$^{-1}$.
This research represents the first detailed investigation into the chemical composition of the cluster, measuring the abundances of a wide range of elements: odd-Z elements (Na, Al), $\alpha$ elements (Mg, Si, Ca, Ti), Fe-peak elements (Sc, V, Cr, Mn, Co, Fe, Ni, Cu, Zn), and $n$-elements (Sr, Y, Ba, La, Ce, Nd).
The cluster shows chemical homogeneity among its members, and the abundances derived from dwarf and giant stars are in excellent agreement. In addition, the chemical pattern drawn by both approaches, \Space\ and SYNTHE, agrees within a precision of 0.1\,dex for most of the elements, further confirming the consistency between both methodologies, as found with atmospheric parameters.
NGC\,2509 exhibits solar-like [X/Fe] ratios that align well with Galactic trends (seen in extensive OC surveys like $Gaia$-ESO, OCCAM, and SPA)
and mild supersolar metallicity ([Fe/H]$\approx$0.1\,dex). This is a value somewhat higher than that observed in most of the OCs found at a similar galactocentric distance but still compatible with the Galactic gradient traced by them.
Our findings lead us to conclude that the chemical composition of NGC\,2509 is fully compatible with that of the Galactic thin disk.

We also investigated the age of NGC\,2509, as the results of the most recent
papers \citep{deJuanOvelar2020,HR2024} showed some discrepancies, resulting
in values in the range 900--1600\,Myr.  We derived the cluster age from the
abundance of Li among MS stars (\teff<6500\,K) and the isochrone-fitting
method, obtaining similar values in the range 1200 - 1300\,Myr.  In
agreement with previous works, we find that the reddening in the cluster
field is small ($A_0\approx0.25$\,mag), probably due to its location towards
the Galactic anticenter.  Unlike other OCs of similar age that exhibit an
eMSTO, in the CMD of NGC\,2509 the MS stars are distributed following a
narrow strip.  This is consistent with the fact that the stars that populate
the MSTO show a narrow distribution of the \vsini\ that cover the range
20--137~\kmsec.

\section*{Data availability}
Table~\ref{t:list} is only available in electronic form at the CDS via anonymous ftp to cdsarc.u-strasbg.fr (130.79.128.5) or via http://cdsweb.u-strasbg.fr/cgi-bin/qcat?J/A+A/.

\begin{acknowledgements}
The data of this work are based on observations collected at the European Southern Observatory under ESO program 112.25DP.001 or 0112.D-2214 and processed data created thereof. This research has made use of the VizieR catalog access tool, CDS,
Strasbourg, France (DOI : 10.26093/cds/vizier).  The original description of
the VizieR service was published in 2000, A\&AS 143, 23.  Use of the NASA's
Astrophysical Data System and TOPCAT \citep{topcat} are also acknowledged. 
We acknowledge funding from Bando Astrofisica Fondamentale INAF
2022 (High-resolution spectroscopy of open clusters, PI Bragaglia). 
We acknowledge funding from Bando Astrofisica Fondamentale INAF 2023 
(Open Clusters and stellar structures in the local Galactic disk, PI A. Vallenari).
BC acknowledges funding from ASI grant n. 2025-10-HH.0.
AF and JAS acknowledge funding from the Large Grant INAF-2024 “Spectral Key features of Young stellar objects: Wind-Accretion LinKs Explored in the infraRed (SKYWALKER)”. VD acknowledges partial support by the INAF Minigrant 2024 MUGS.

\end{acknowledgements}

  \bibliographystyle{aa}
  \bibliography{biblio.bib}

\newpage\eject
\begin{appendix}\label{AppA}
\section{Normalization routine}\label{sec_norm}
GIRAFFE and UVES spectra have features that make normalization made with
IRAF packages very costly in terms of working time.  In fact, some pixels at
the beginning and the end of the spectrum have flux zero; besides narrow
high spikes are present in some spectra caused by cosmic rays.  Such
unwanted features cause the IRAF {\it continuum} package to perform a
sub-optimal work in normalizing the flux unless such defects are removed
with a careful cosmetic action of handling the spectral intervals one by
one.  To avoid such time-consuming work we wrote an ad hoc normalization
routine in python for an automated spectra cleaning.  The algorithm can be
outlined as follow:

\begin{enumerate}
\item Clip out a fixed number of pixels at the beginning and at the end of the spectrum (quantity optimized as a function of the spectrum type).
\item Identify the cosmic ray spikes by using the python function {\it scipy.signal.find\_peaks} \citep{scipy} and flatten them out with a linear interpolating function obtaining the flux $f_{ini}$.
\item Convolve the spectrum $f_{ini}$ with a Gaussian kernel (which width can be different for different resolutions and chosen as it
suits) obtaining the flux $f_c$.
\item Obtain the first normalized flux $f_n$ by dividing the spectrum $f_{ini}$ by $f_c$.
\item identify the pixels in $f_n$ that are i) larger than the median by 3$\sigma$ ii) smaller than the median by 1$\sigma$. For these pixels the flux of $f_{n}$ is flatten at the given $\sigma$ level.
\item Convolve the flux $f_n$ obtained in step 5 with a Gaussian kernel obtaining the flux $f_c$.
\item obtain a new normalized flux $f_n$ by dividing $f_{ini}$ by the convolved flux $f_c$. 
\item repeat steps 5, 6, and 7 10 times.
\end{enumerate}

This algorithm allows fo an automated and uniform spectra normalization
without manual intervention.

\section{The S$\Delta_{color}$ parameter}\label{sec_vert_color}

\begin{figure}[]
\centering
\includegraphics[width=9cm]{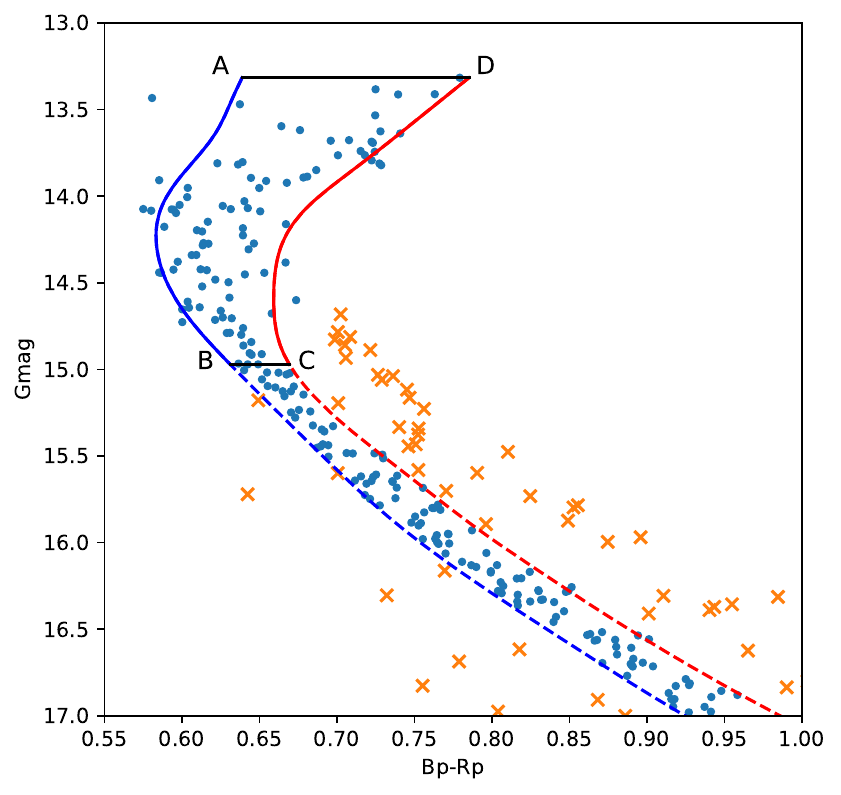}
\caption{MS of for NGC~2509 with blue and red fiducial lines traced as described in the text. The dashed lines trace the MS while the solid lines ABCD frame the TO. Orange crosses are the stars rejected before tracing the fiducial lines. The segment BC represents the base of the TO.}
\label{f:MSTO_DeltaColor}
\end{figure}

This parameter was used by \citet{cordoni24} and originally proposed by \citet{milone2017}. It is derived by first tracing two fiducial lines that limit the blue and red borders of the MSTO (see Fig.~\ref{f:MSTO_DeltaColor}) and defining the verticalized color for each star as
$$
\Delta_{color} = w_{TO} \frac{C - C_{blue}}{C_{red} - C_{blue}},
$$
where $C_{blue}$ and $C_{red}$ are the colors (Gaia $Bp-Rp$ in this case) of
the two fiducial lines at the stars's  magnitude $G$, $C$ is the color of the
star, and $w_{TO}$ is the color width of the base of the TO.  This is the
width of the line between the points B and C in Fig.\ref{f:MSTO_DeltaColor}
chosen at $G_{mag}=15$.  The points A,D,B,C in the same figure, define the
turnoff region for the determination of $\Delta_{color}$.  The fiducial
lines are selected with the following procedure:

\begin{enumerate}
\item For every star with magnitude $G$ we consider all stars in the magnitude interval $\Delta G = G\pm0.1$ and compute the median color $Bp-Rp$ and its standard deviation $\sigma$. 
\item We reject the stars that have color beyond $1\sigma$ from the median (orange crosses in Fig.~\ref{f:MSTO_DeltaColor}.)
\item For the left stars, we compute the median color ($BpRp_{median}$), the blue and red borders as the 0.2\% and 99.8\% quantiles ($\pm3\sigma$) of the distribution for each $\Delta G$ at the $G$ of each star
\item Trace two smoothing splines \citep[we used the python function {\it make\_smoothing\_spline} in {\it scipy.interpolate}][]{scipy} interpolating the values of the blue and red borders. The interpolating lines correspond to the blue and red lines in Fig.\ref{f:MSTO_DeltaColor}.
\end{enumerate}

\noindent We define $S\Delta_{color}$ as the color width between the quantiles 16\% and 84\% of the $\Delta_{color}$ distribution.

\onecolumn
\section{Auxiliary figures and tables}
In this appendix we moved figures and tables referred in the text in order to make the text more readable.

%\centering
\scriptsize{\setlength{\tabcolsep}{1mm}
\setlength{\tabcolsep}{1mm}
\setlength{\LTcapwidth}{\textwidth}
\begin{table*}[!ht]
\tiny
\centering
\caption{Targets, observing modes, radial velocity, vsini, and AP (measured by ROTFIT) (extract). }
\label{t:list}
\begin{tabular}{llllrrrrrrrrrrrrlll}
\hline\hline
$Gaia$-DR3 ID        &1of3 &2of3 &3of3 &  G    & G$_{BP}$-G$_{RP}$    
 &Teff &eTeff &logg &elogg & [Fe/H] &eFeH &vsini &evsini &RV &eRV &Sp.Type & HR used\\
 \hline\hline
  5714216325322800128 & Y & N & Y & 13.037354 & 1.187693    & 4840 & 58   & 2.84 & 0.22  &    0.04  & 0.08 &   0.2 & 0.5  & 60.98 & 0.11  &  K0III   & U580\\
  5714215638128211968 & Y & N & N & 12.897546 & 1.226348    & 4694 & 80   & 2.67 & 0.15  &    0.14  & 0.08 &   0.2 & 0.5  & 60.99 & 0.10  &  K1.5III & U580\\
  5714216840718856064 & Y & N & N & 12.947951 & 1.204591    & 4742 & 104  & 2.70 & 0.19  &    0.04  & 0.08 &   1.0 & 1.3  & 61.15 & 0.08  &  G8III   & U580  \\
  5714215947365908352 & Y & Y & N & 13.072523 & 1.205736    & 4768 & 100  & 2.75 & 0.19  &    0.07  & 0.09 &   0.2 & 0.5  & 60.93 & 0.07  &  K0III   & U580 \\
  5714218833583650304 & Y & Y & N & 12.877121 & 1.199892    & 4840 & 58   & 2.75 & 0.17  &    0.06  & 0.09 &   1.1 & 1.4  & 59.11 & 0.09  &  G8III   & U580 \\
  5714209689591479424 & Y & N & N & 13.075513 & 1.206331    & 4737 & 101  & 2.67 & 0.18  &    0.07  & 0.09 &   1.4 & 1.5  & 60.40 & 0.11  &  G8III   & U580 \\
  5714218283827858304 & Y & N & N & 13.148229 & 1.200976    & 4992 & 113  & 2.98 & 0.23  &    0.05  & 0.08 &   1.9 & 1.5  & 61.34 & 0.12  &  G8III   & U580 \\
  5714216737639647360 & N & Y & N & 12.989202 & 1.239747    & 4704 & 83   & 2.76 & 0.19  &    0.10  & 0.09 &   0.3 & 0.7  & 61.11 & 0.11  &  K1.5III & U580 \\
  5714215191451661440 & N & Y & N & 12.915168 & 1.209816    & 4727 & 101  & 2.70 & 0.19  &    0.05  & 0.09 &   0.2 & 0.5  & 61.58 & 0.10  &  K1.5III & U580 \\
  5714216565840947328 & N & Y & N & 12.79095  & 1.223884    & 4690 & 80   & 2.65 & 0.16  &    0.00  & 0.10 &   1.0 & 1.3  & 60.38 & 0.10  &  K1.5III & U580 \\
\end{tabular}
\tablefoot{The full version of this table is available at the CDS. Most of the columns are self-explanatory. The columns 1of3, 2of3, and 3of3 indicate the observing runs as described in Table~\ref{t:logobs}. For UVES they indicate whether UVES spectra were gathered (Yes or No). For GIRAFFE they indicate which HR interval was observed. For the sake of shortness we only report the first 10 lines and 18 columns. }
\end{table*}
}

\scriptsize{\setlength{\tabcolsep}{1mm}
\begin{table*}[!hb]
\tiny
\centering
    \caption{Stellar parameters and chemical abundances derived with \Space\ (extract). }\label{t:TGM_ABD_SPACE}
    \begin{tabular}{lrrrrrrrrrrrrrrrrrrrc}
\hline\hline
$Gaia$-DR3 ID & conv & RVb & RVr & FWHMb & FWHMr & S/N & chisq & Teff & T\_l & T\_h & logg & L\_l & L\_h & MH & MH\_l & MH\_h & Fe & Fe\_l & Fe\_h & ...\\
\hline
5714213022486184192 & 0 & 1.10 & 1.80 & 0.44 & 0.46 & 26 & 1.39 & 5844 & 5830 & 6012 & 4.11 & 3.98 & 4.35 & 0.19 & 0.18 & 0.27 & 0.15 & 0.13 & 0.21 & ...\\
5714408804275744000 & 0 & 1.60 & -0.80 & 0.47 & 0.60 & 32 & 1.48 & 5991 & 5938 & 6044 & 4.09 & 3.98 & 4.19 & 0.12 & 0.10 & 0.15 & 0.17 & 0.14 & 0.18 & ...\\
5714208658799247744 & 0 & 2.20 & 1.30 & 0.48 & 0.58 & 39 & 1.35 & 6144 & 6107 & 6189 & 3.98 & 3.89 & 4.04 & 0.25 & 0.23 & 0.27 & 0.30 & 0.28 & 0.32 & ...\\
5714214779134817536 & 0 & 1.50 & 1.00 & 0.41 & 0.53 & 23 & 1.46 & 5866 & 5827 &  & 4.15 & 4.07 &  & 0.38 & 0.36 &  & 0.44 & 0.42 & 0.47 & ...\\
5714208731820904576 & 0 & 1.60 & -0.00 & 0.44 & 0.55 & 20 & 1.53 & 5695 & 5642 & 5777 & 3.86 & 3.73 & 3.98 & 0.16 & 0.12 & 0.20 & 0.21 & 0.18 & 0.24 & ...\\
5714208525662494976 & 0 & 1.50 & 1.40 & 0.42 & 0.51 & 23 & 1.41 & 5735 & 5696 & 5806 & 4.15 & 4.08 & 4.30 & 0.34 & 0.31 & 0.37 & 0.35 & 0.32 & 0.37 & ...\\
5714204917889795968 & 0 & 0.70 & 1.00 & 0.39 & 0.52 & 22 & 1.47 & 5685 & 5664 & 5847 & 3.98 & 3.88 & 4.21 & 0.26 & 0.24 & 0.34 & 0.26 & 0.22 & 0.29 & ...\\
5714222853673053952 & 0 & 1.40 & 1.40 & 0.39 & 0.52 & 27 & 1.41 & 5761 & 5736 & 5857 & 4.24 & 4.18 & 4.42 & 0.22 & 0.21 & 0.28 & 0.22 & 0.21 & 0.26 & ...\\
5714408980373914496 & 0 & 1.70 & 1.70 & 0.42 & 0.59 & 25 & 1.41 & 5968 & 5936 & 6070 & 4.10 & 4.03 & 4.26 & 0.25 & 0.24 & 0.30 & 0.31 & 0.28 & 0.33 & ...\\
5714216084804886272 & 0 & 1.20 & 1.00 & 0.40 & 0.50 & 29 & 1.54 & 5865 & 5832 & 5925 & 4.05 & 3.97 & 4.15 & 0.18 & 0.16 & 0.21 & 0.19 & 0.17 & 0.21 & ...\\
5714222956752259456 & 0 & 1.10 & 2.40 & 0.44 & 0.70 & 27 & 1.43 & 5824 & 5783 & 5884 & 4.04 & 3.96 & 4.16 & 0.26 & 0.24 & 0.29 & 0.27 & 0.24 & 0.29 & ...\\
5714220379771938048 & 0 & 1.90 & 1.30 & 0.45 & 0.50 & 32 & 1.43 & 5960 & 5929 & 6017 & 4.05 & 3.98 & 4.15 & 0.22 & 0.20 & 0.25 & 0.25 & 0.23 & 0.27 & ...\\
5714207804108026240 & 0 & 1.70 & 1.20 & 0.41 & 0.50 & 27 & 1.43 & 5918 & 5892 & 6004 & 4.09 & 4.01 & 4.24 & 0.30 & 0.28 & 0.34 & 0.30 & 0.27 & 0.32 & ...\\
5714213331723824000 & 0 & 1.00 & 0.60 & 0.45 & 0.49 & 28 & 1.45 & 5838 & 5783 & 5897 & 4.00 & 3.90 & 4.11 & 0.18 & 0.15 & 0.21 & 0.20 & 0.16 & 0.22 & ...\\
5714217734072478976 & 0 & 3.00 & 1.50 & 0.48 & 0.56 & 37 & 1.39 & 5973 & 5918 & 6035 & 3.89 & 3.79 & 4.03 & 0.24 & 0.21 & 0.27 & 0.14 & 0.12 & 0.17 & ...\\
5714219245900567936 & 0 & 1.50 & 1.70 & 0.45 & 0.53 & 43 & 1.48 & 6202 & 6156 & 6229 & 4.05 & 3.97 & 4.11 & 0.10 & 0.08 & 0.11 & 0.10 & 0.08 & 0.11 & ...\\
5714221444923778944 & 0 & 0.90 & 1.10 & 0.35 & 0.51 & 39 & 1.47 & 6403 & 6341 & 6472 & 4.17 & 4.05 & 4.24 & 0.12 & 0.09 & 0.14 & 0.19 & 0.16 & 0.21 & ...\\
5714215977425382656 & 0 & 1.50 & 0.40 & 0.37 & 0.41 & 20 & 1.61 & 5771 & 5732 & 5933 & 4.86 & 4.82 & 5.12 & 0.26 & 0.24 & 0.33 & 0.28 & 0.25 & 0.32 & ...\\
5714215019652973056 & 0 & 1.50 & 1.10 & 0.42 & 0.49 & 62 & 1.36 & 6204 & 6180 & 6242 & 4.10 & 4.05 & 4.16 & 0.07 & 0.06 & 0.09 & 0.09 & 0.08 & 0.10 & ...\\
5714215908705903360 & 0 & 1.60 & 1.80 & 0.44 & 0.55 & 31 & 1.45 & 6032 & 5989 & 6090 & 4.07 & 3.98 & 4.17 & 0.25 & 0.23 & 0.28 & 0.27 & 0.25 & 0.29 & ...\\

\hline
\end{tabular}
\tablefoot{For the sake of brevity, we report only the first 20 stars and the first 20 columns. 
The full table and explanation of the columns are available at the CDS. The table is pruned from SB2 stars and from stars with \vsini\ $>10$ \kmsec. 
The columns report (from left to right): Gaia source ID; \Space\ convergence (conv); correction of the radial velocity for the blue and red part of the spectrum (RVb and RVr); Full Width Half Maximum estimated for the blue and the red part of the spectrum (FWHMb and FWHMr); Signal-to-Noise estimated by \Space\ (S/N); $\chi^2$ between the observed spectrum and the model (chisq); effective temperature (Teff); upper and lower limits holding 68\% of probability of the Teff measurement (T\_l and T\_h); similarly for gravity (logg), metallicity (MH) and all the elements measured starting with Fe.}    
\end{table*}
}
\begin{figure*}[htb]
\includegraphics[width=6cm]{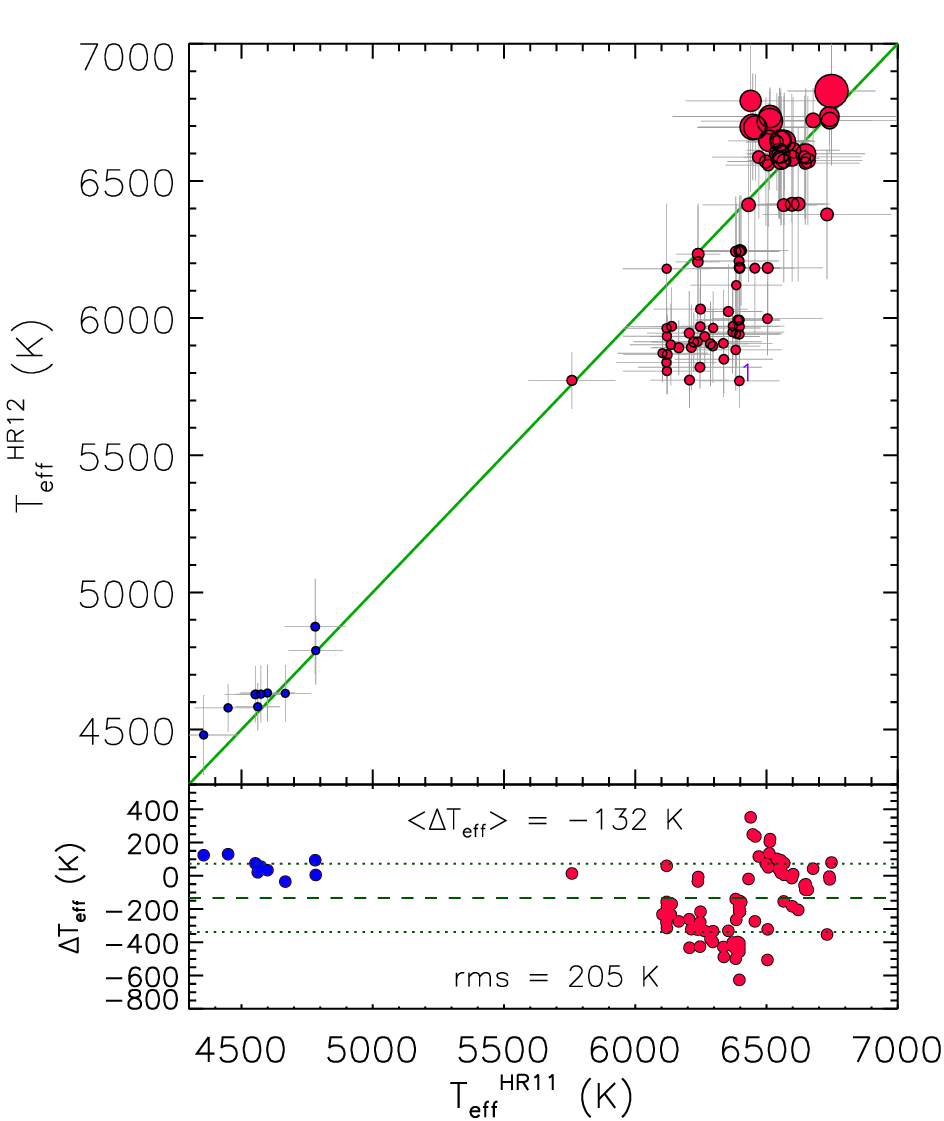}
\includegraphics[width=6cm]{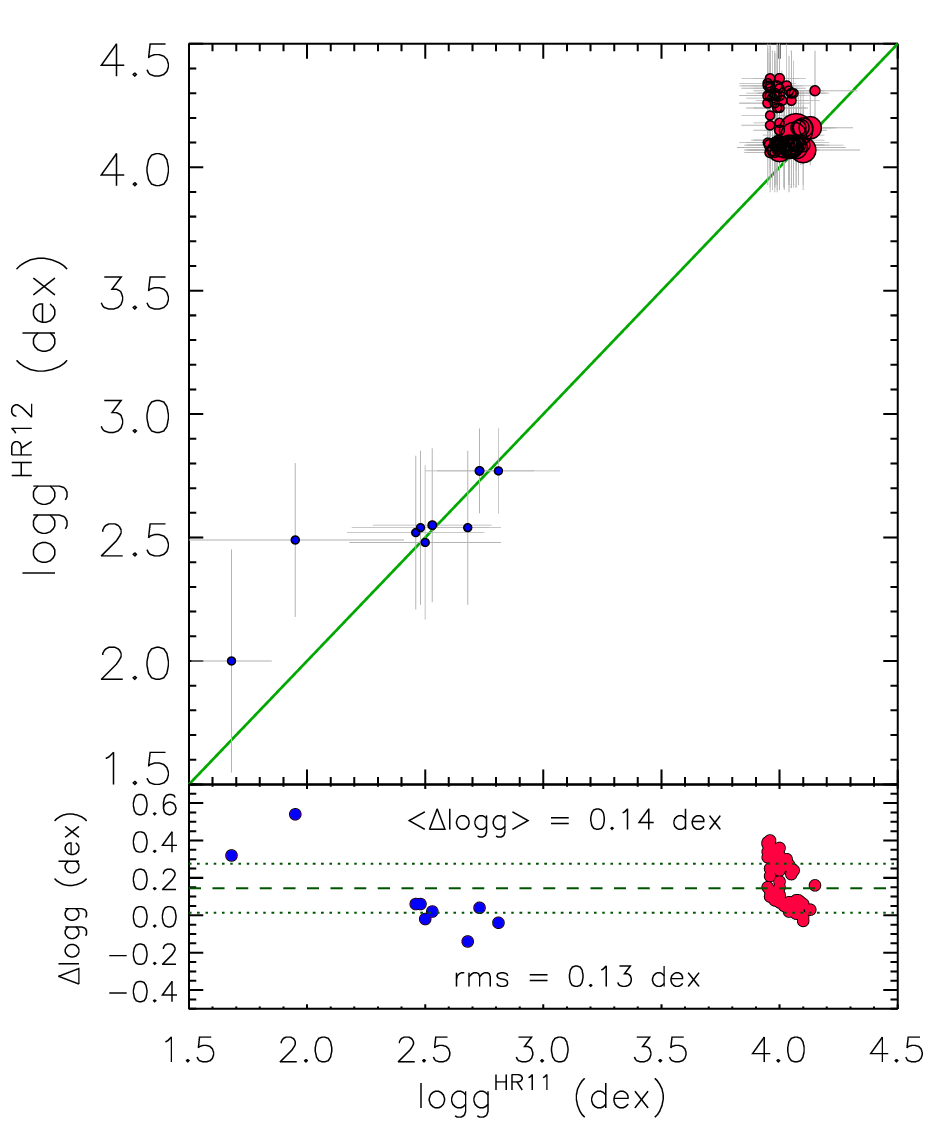}
\includegraphics[width=6cm]{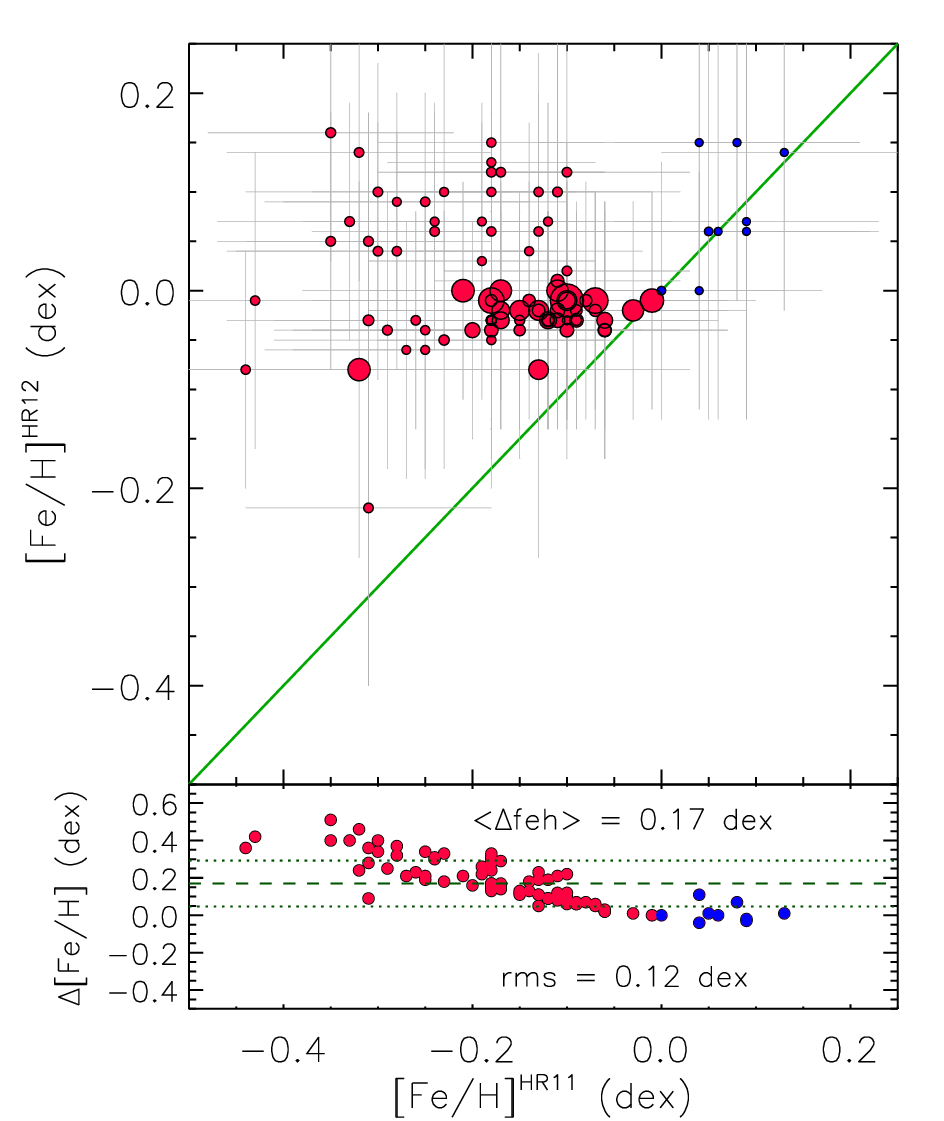}
\caption{Comparison between \teff, \logg, and \feh\ (left to
right) measured from the HR11 and HR12 GIRAFFE spectra by the \rotfit\ software.  Red and blue
dots are used for main-sequence and giant stars, respectively.  In the upper
boxes, the symbol size scales with \vsini.  The one-to-one relation is shown
by the full green line in each of the upper boxes.  The differences are
displayed in the bottom panels along with the average value  (dashed lines)
and the standard deviations (green dot-dashed lines).}
\label{Fig:APs}
\end{figure*}

\begin{figure*}
%\centering
\hspace{-.4cm}
\includegraphics[scale=0.35]{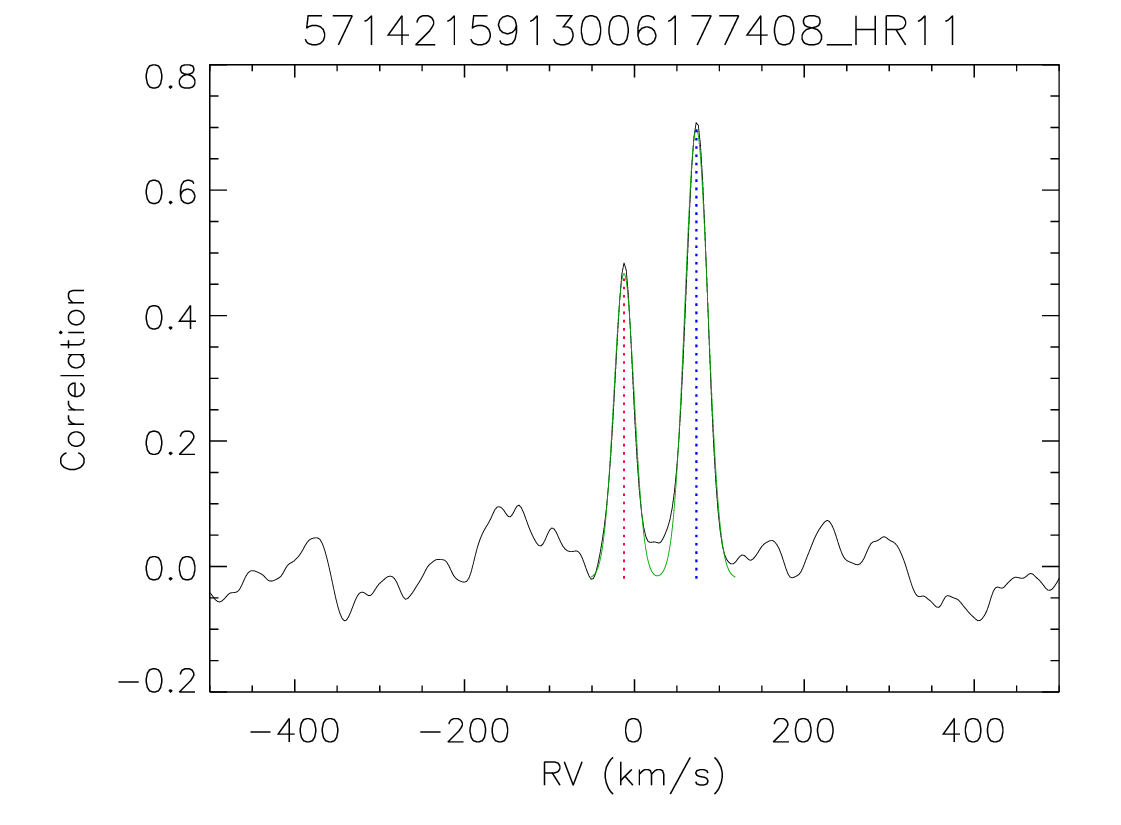}
\hspace{-.6cm}
\includegraphics[scale=0.35]{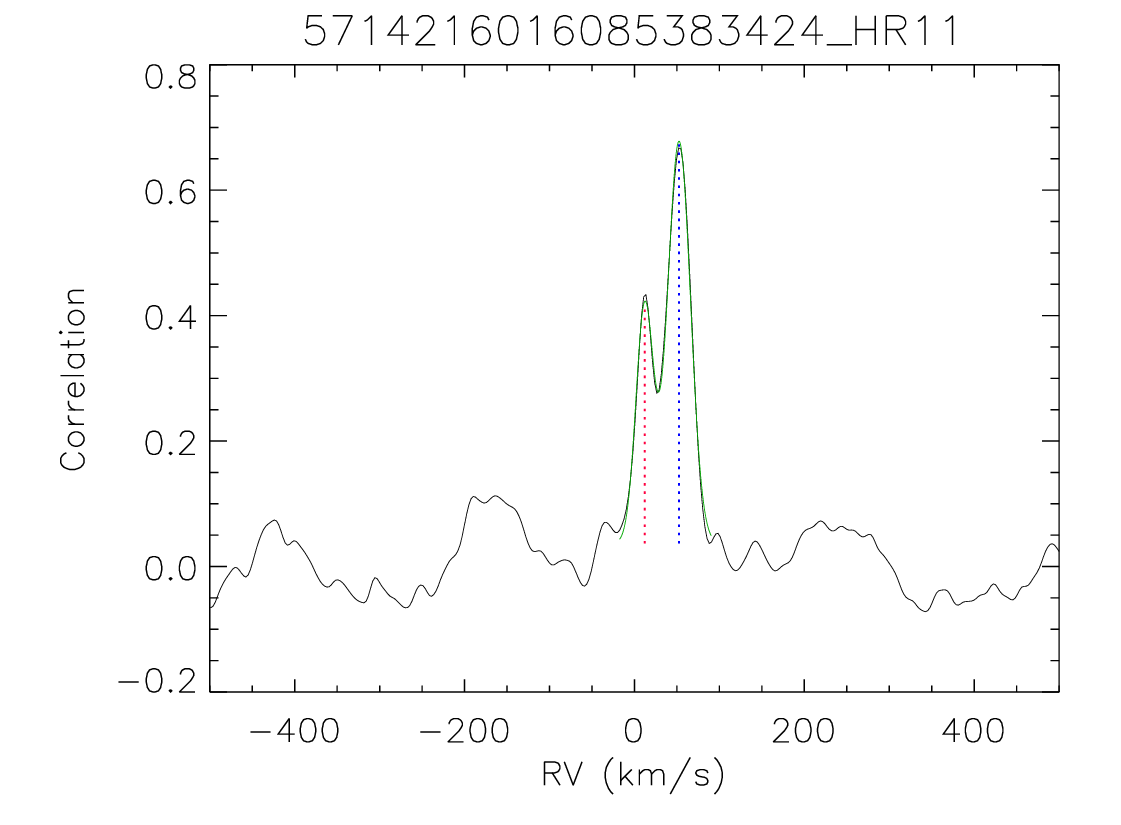}
\hspace{-.6cm}
\includegraphics[scale=0.35]{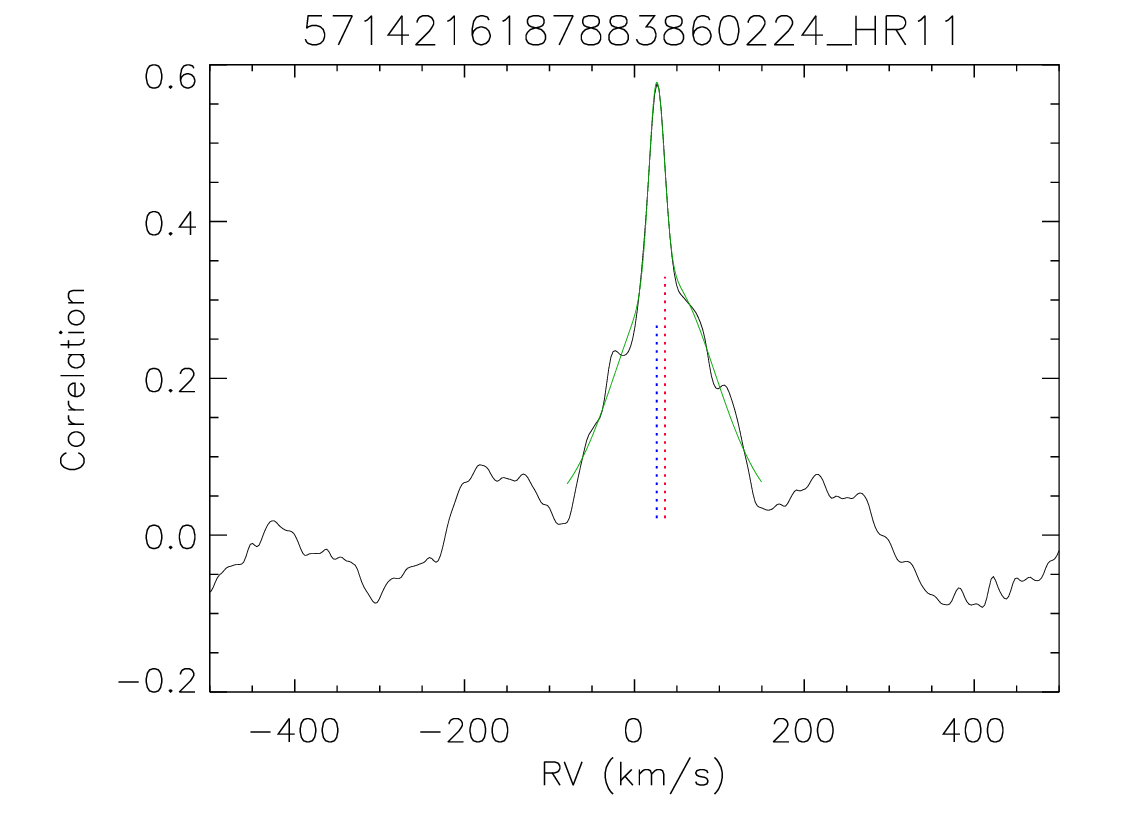}
\caption{Example of cross-correlation functions of double-lined
binaries (black lines).  The two-Gaussian fit to the CCF is overlaid with a
green line.  The radial velocity of the primary and secondary component  are
marked with a blue a red dotted line, respectively.  }
\label{Fig:SB2}
\end{figure*}

\begin{table}[ht]
  \caption{Radial velocity of the primary ($RV_1$) and the secondary ($RV_2$) components of the double-lined systems discovered in the present work.}
\begin{center}
\begin{tabular}{ccrrll}
\hline\hline
\noalign{\smallskip}
         \gaia-DR3 ID &      HJD\tablefootmark{a}                 &      $RV_1$  &  $RV_2$    &  Setup & Remarks  \\ 
                      &  \scriptsize{(2\,460\,000+)} &   (\kms)  &  (\kms)    &          &   \\ 
\noalign{\smallskip}
\hline
\noalign{\smallskip}
  5714210003122616960  & 297.74392 &  74.11$\pm$14.21 &  6.01$\pm$7.55     & HR15N  & Complex CCF \\
  "   "                             & 332.68484 &  -0.87$\pm$2.37  & 71.74$\pm$5.55     & HR11   &  \\ 
  "   "                             & 378.64123 & 110.22$\pm$4.18  & 26.97$\pm$5.83     & HR12   &  \\ 
  5714215913006177408  & 297.74392 &  59.22$\pm$4.25  & 59.22$\pm$4.25\tablefootmark{b} & HR15N  & Conjunction \\
  "   "                             & 332.68484 &  97.88$\pm$0.28  & 12.67$\pm$0.39     & HR11   &  \\ 
  "   "                             & 378.64123 &  10.24$\pm$0.20  & 116.13$\pm$0.27    & HR12   &  \\   
  5714216016085383424  & 297.74392 &  61.79$\pm$3.69  & 61.79$\pm$3.69$^b$ & HR15N  & Conjunction \\ 
  "   "                             & 332.68484 &  77.51$\pm$0.72  & 37.14$\pm$1.09     & HR11   &  \\ 
  "   "                             & 378.64123 &  37.89$\pm$0.26  & 84.98$\pm$0.31     & HR12   &  \\
  5714216050444885632  & 297.74392 & 102.06$\pm$0.67  & 63.24$\pm$1.08     & HR15N  &  Triple? \\
  "   "                             & 332.68484 &  80.74$\pm$1.38  & 58.94$\pm$5.43     & HR11   &  \\ 
  "   "                             & 378.64123 &  30.25$\pm$0.46  & 59.08$\pm$1.42     & HR12   &  \\
  5714216187883860224  & 297.74392 &  50.58$\pm$0.78  & 61.48$\pm$2.06     & HR15N  & Broad+narrow \\
  "   "                             & 332.68484 &  51.08$\pm$0.62  & 61.24$\pm$1.33     & HR11   & "   " \\  
  "   "                             & 378.64123 &  50.85$\pm$0.49  & 63.78$\pm$1.24     & HR12   & "   " \\
  5714216256593408128  & 297.74392 &  33.86$\pm$0.97  & 77.24$\pm$2.04     & HR15N  &  \\
  "   "                             & 332.68484 &  59.74$\pm$1.93  &  0.80$\pm$2.03     & HR11   &  \\ 
  5714216359682737536  & 297.74392 &  59.56$\pm$0.41  & 66.57$\pm$1.21     & HR15N  & Broad+narrow \\
  "   "                             & 332.68484 &  58.65$\pm$0.58  & 58.40$\pm$1.19     & HR11   & "   " \\ 
  "   "                             & 378.64123 &  59.87$\pm$0.39  & 61.74$\pm$1.00     & HR12   & "   " \\ 
\noalign{\smallskip}
\hline
\end{tabular}
\end{center}
\tablefoot{
\tablefoottext{a}{HJD is the heliocentric Julian date at mid exposure.}
\tablefoottext{b}{Blended CCF peaks observed near a conjunction.}
}
\label{Tab:SB2}
\end{table}

\newpage

\begin{table}[ht]
  \caption{Stars with variable radial velocity, candidates for SB1 systems.}
\begin{center}
\begin{tabular}{ccrll}
\hline\hline
\noalign{\smallskip}
         \gaia-DR3 ID &      HJD\tablefootmark{a}                 &      $RV$    &  Setup & Remarks  \\ 
                      &  \scriptsize{(2\,460\,000+)} &   (\kms)  &          &   \\ 
\noalign{\smallskip}
\hline
\noalign{\smallskip}
 5714208663101422464  & 297.74392 & 50.22$\pm$2.06 &  HR15N  & SB2? Slightly asymmetric CCF. \\ 
  "   "                            & 332.68484 & 58.65$\pm$3.39 &  HR11   & "  " \\  
  "   "                            & 378.64123 & 63.22$\pm$1.80 &  HR12   & "  " \\ 
 5714216325322798208  & 297.74392 & 73.86$\pm$1.42 &  HR15N  &  SB1? Broad CCF. \\  
  "   "                            & 332.68484 & 55.33$\pm$2.94 &  HR11   & "  " \\
  "   "                            & 378.64123 & 66.54$\pm$1.30 &  HR12   & "  " \\
 5714216325322800512 & 297.74392 & 60.49$\pm$0.70 &  HR15N  &  SB1 \\ 
  "   "                            & 332.68484 & 67.10$\pm$0.60 &  HR11   & "  " \\
  "   "                            & 378.64123 & 73.05$\pm$0.42 &  HR12   & "  " \\
 5714220070534267904  & 297.74392 & 45.53$\pm$0.30 &  HR15N  &  SB1 \\    
  "   "                            & 378.64123 & 81.32$\pm$0.24 &  HR12   & "  " \\  
 5714216806359113216  & 297.74392 & 78.14$\pm$0.42 &  HR15N  &  SB1 \\    
  "   "                            & 332.68484 & 45.66$\pm$0.55 &  HR11   & "  " \\    
 5714207735388551936 & 297.74392 & 63.66$\pm$1.04 &  HR15N  &  SB1? Broad CCF. \\   
  "   "                            & 332.68484 & 69.89$\pm$5.48 &  HR11   & "  " \\  
  "   "                            & 378.64123 & 60.73$\pm$3.44 &  HR12   & "  " \\  
 5714218318187587072  & 297.74392 & 62.48$\pm$0.60 &  HR15N  & SB2? Slightly asymmetric CCF. \\       
  "   "                            & 332.68484 & 63.33$\pm$2.16 &  HR11   & "  " \\                              
  "   "                            & 378.64123 & 68.68$\pm$2.42 &  HR12   & "  " \\                             
\noalign{\smallskip}
\hline
\end{tabular}
\end{center}
\tablefoot{
\tablefoottext{a}{HJD is the heliocentric Julian date at mid exposure.}
}
\label{Tab:SB1}
\end{table}

\newpage

\renewcommand{\tabcolsep}{0.1cm}

\begin{figure*}[]
\centering
\includegraphics[width=16cm]{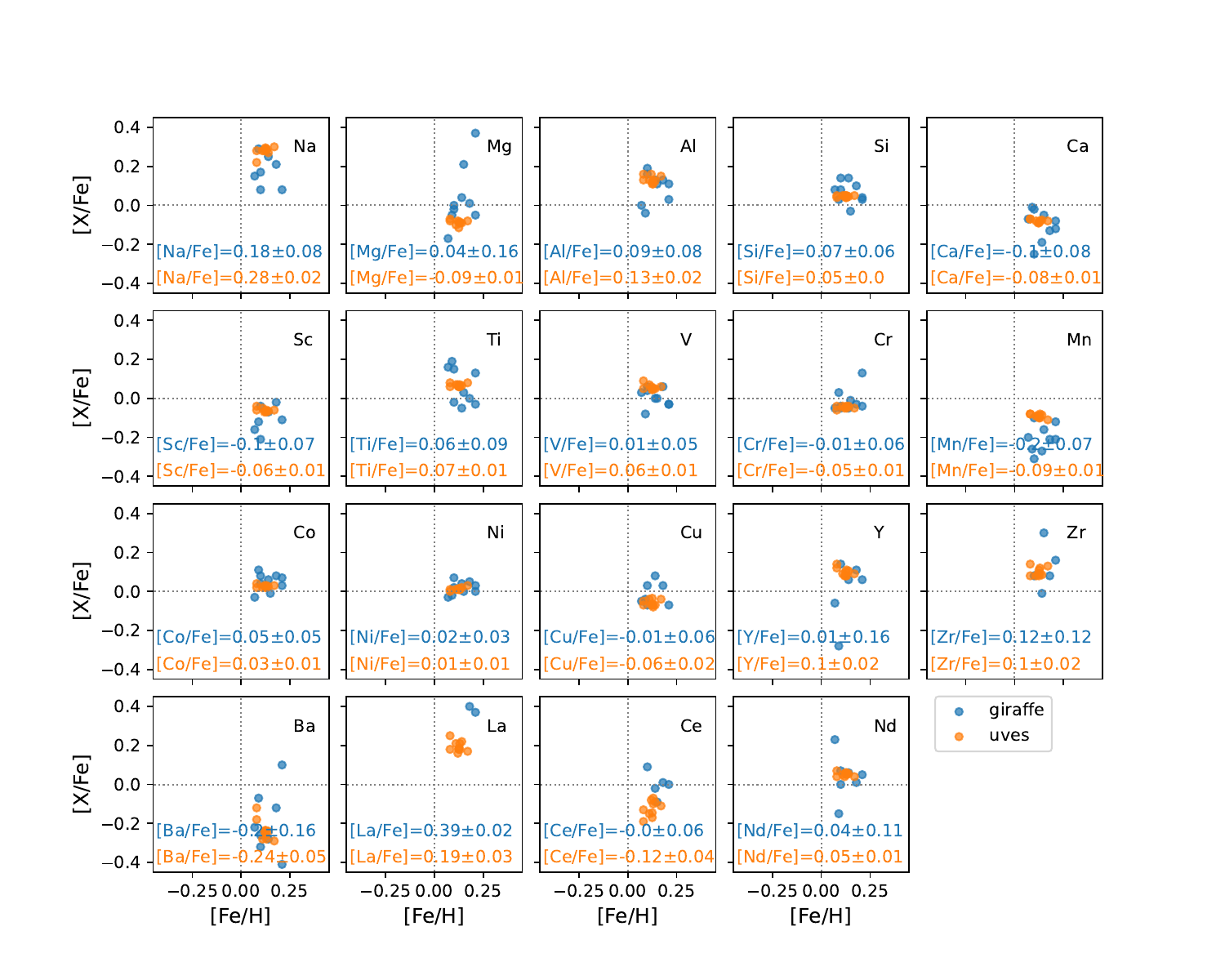}
\caption{Results from \Space: relative abundances [X/Fe] versus [Fe/H] of GIRAFFE (blues points)
and UVES (orange points) spectra selected as described in the text. The blue
and orange texts report the average abundances and standard deviations for each
element of the cluster computed with GIRAFFE and UVES spectra, respectively.
}
\label{f:ABD_SPACE}
\end{figure*}

\begin{figure*}[htb]
\includegraphics[width=6.3cm]{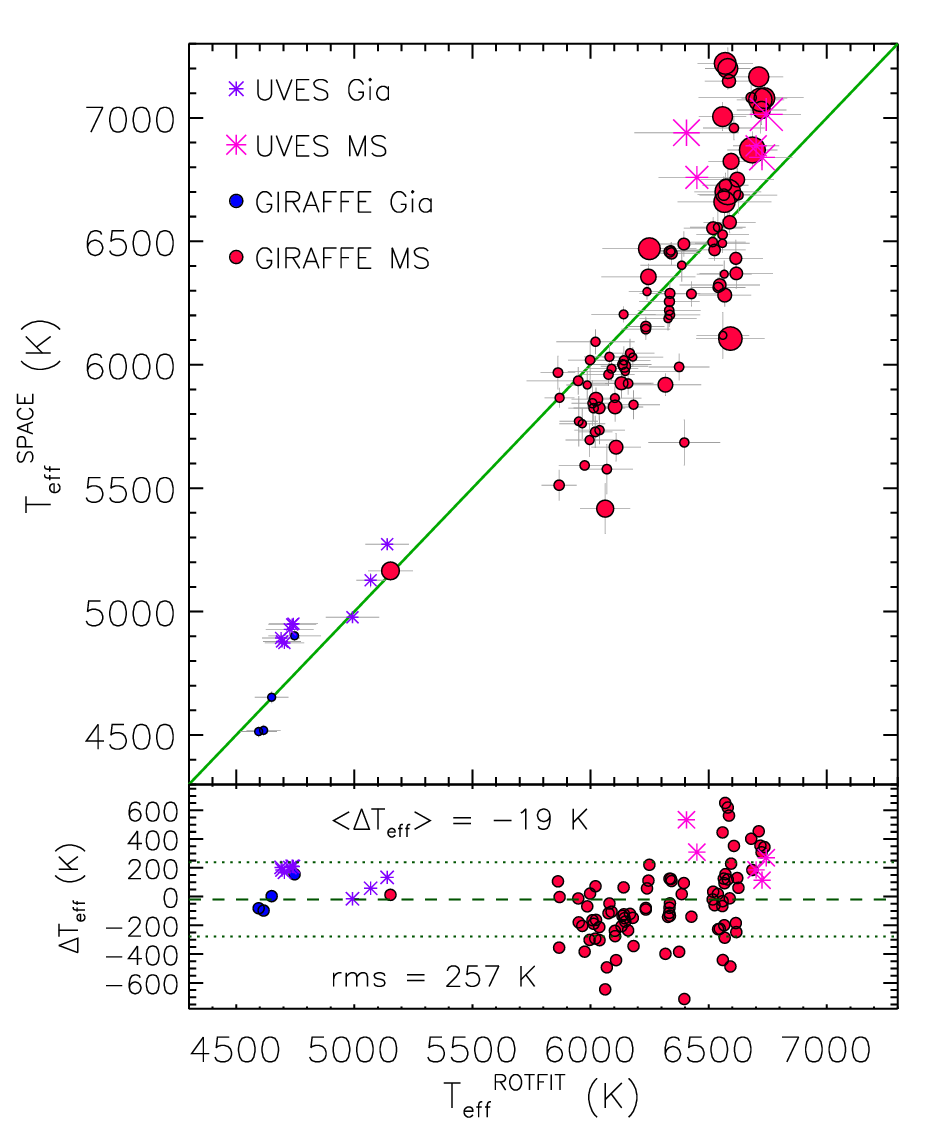}
\hspace{-.3cm}
\includegraphics[width=6.3cm]{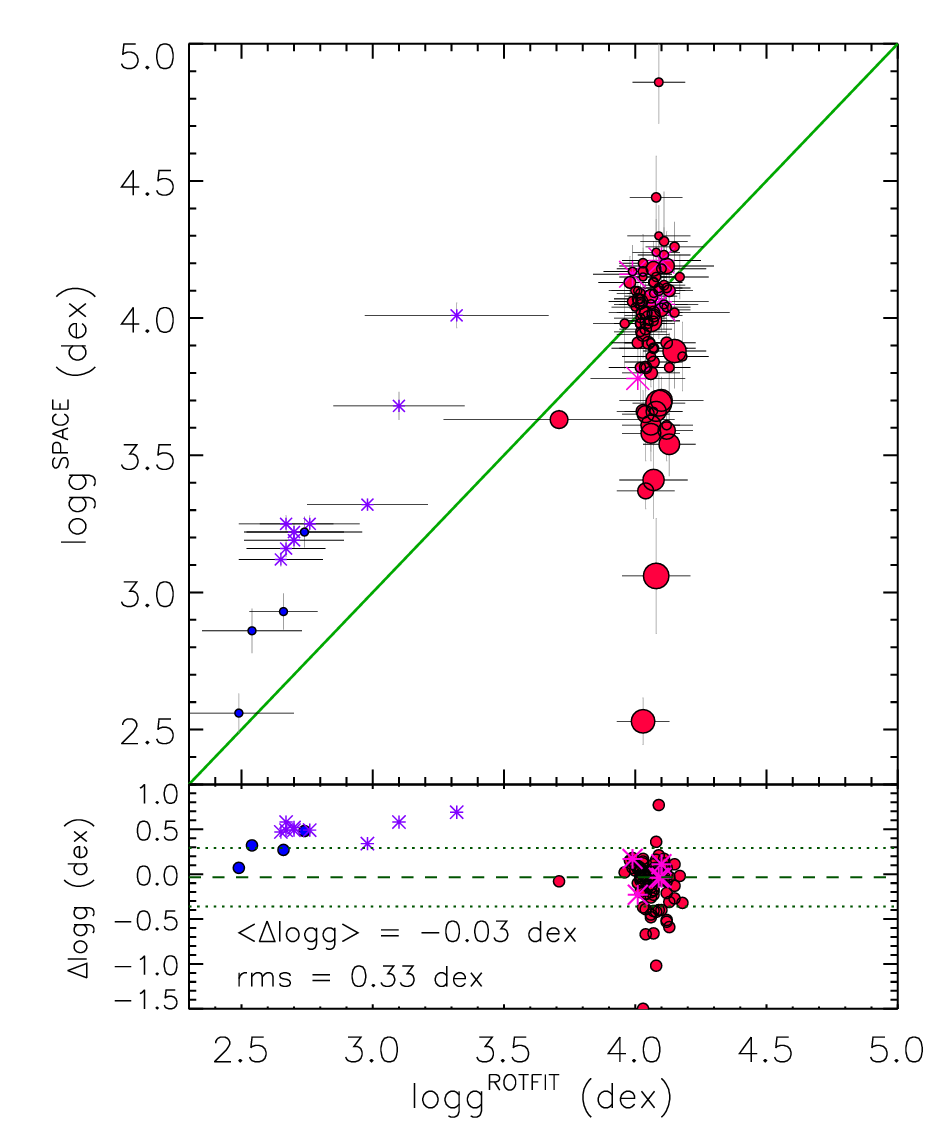}
\hspace{-.3cm}
\includegraphics[width=6.3cm]{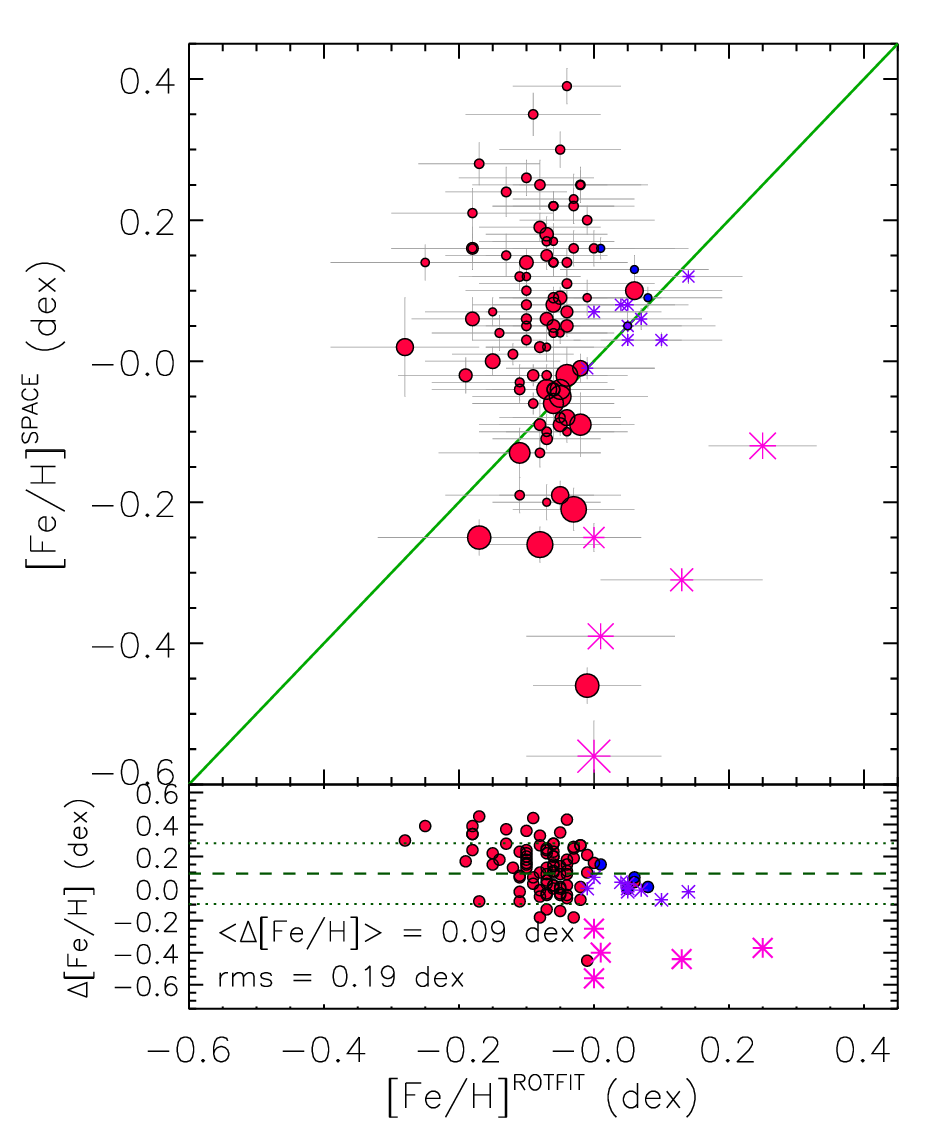}
\caption{Comparison of the atmospheric parameters (\teff, \logg, and \feh,
from left to right) derived with \Space\ and \rotfit.  Different
symbols are used for main-sequence and giant stars, distinguishing also
between UVES and GIRAFFE spectra.  In the upper boxes, the symbol size
scales with \vsini\ and the full green line represents the one-to-one
relationship.  The lower boxes display the difference of the parameters
(\Space-\rotfit) and report the mean differences ($<\Delta$\teff$>$,
$<\Delta$\logg$>$, and $<\Delta$\feh$>$) and standard deviations (rms),
which are also marked by horizontal dashed and dotted lines, respectively.
}
\label{Fig:SPACE_ROTFIT}
\end{figure*}

\begin{table}
\caption{Lithium line ($\lambda=6807$\AA) equivalent widths in MS stars below 6500\,K.}\label{tab_eqw}
\begin{center}
\begin{tabular}{lc|cc}
\hline\hline
$Gaia$-DR3 ID  &   $EW_{\textrm{Li}}$ (m\AA)  & $Gaia$-DR3 ID  &   $EW_{\textrm{Li}}$ (m\AA)\\  
\hline
5714208834900087040  &  28.2 $\pm$ 6.4  & ~ 5714217321755349888  &  38.0 $\pm$ 7.1  \\
5714219692577149824  &   4.9 $\pm$ 4.9  & ~ 5714202375269186176  &  64.1 $\pm$ 4.0  \\
5714408980373914496  &  54.3 $\pm$ 7.2  & ~ 5714216634560438272  &  52.6 $\pm$ 4.9  \\
5714220036174536064  &  19.0 $\pm$12.2  & ~ 5714215019652973056  &  31.3 $\pm$ 2.9  \\
5714214779134817536  &  24.7 $\pm$ 6.1  & ~ 5714213645263253760  &  48.5 $\pm$ 4.9  \\
5714216561536744960  &  52.5 $\pm$ 9.0  & ~ 5714218180748805888  &  56.2 $\pm$ 5.6  \\
5714209968771469568  &  66.9 $\pm$ 5.9  & ~ 5714217734072478976  &  81.8 $\pm$ 4.0  \\
5714220070534267904  &  82.5 $\pm$ 5.0  & ~ 5714208933677173248  &  45.0 $\pm$ 3.8  \\
5714215977425382656  &  29.0 $\pm$ 7.2  & ~ 5714216011783739648  &  47.0 $\pm$ 4.2  \\
5714222853673053952  &  42.4 $\pm$ 4.7  & ~ 5714215500689239552  &  61.1 $\pm$ 4.6  \\
5714218489986276992  &  23.4 $\pm$ 5.0  & ~ 5714213331723824000  &  55.9 $\pm$ 7.0  \\
5714207804108026240  &  97.2 $\pm$ 5.6  & ~ 5714208658799247744  &  55.1 $\pm$ 5.6  \\
5714208731820904576  &   7.8 $\pm$ 7.8  & ~ 5714218318187590144  &  12.2 $\pm$ 2.9  \\
5714220792088783488  &  24.2 $\pm$ 6.4  & ~ 5714216256603316736  &   9.3 $\pm$ 2.8  \\
5714213022486184192  &  45.4 $\pm$ 6.6  & ~ 5714216497121665408  &  36.1 $\pm$ 5.0  \\
5714222956752259456  &  27.6 $\pm$ 5.2  & ~ 5714214916573748864  &  88.5 $\pm$ 3.3  \\
5714597611035070848  &  16.5 $\pm$ 7.2  & ~ 5714203573555554048  &  63.2 $\pm$ 4.2  \\
5714213164226900864  &  73.8 $\pm$ 4.0  & ~ 5714221135686174208  &  44.2 $\pm$ 3.8  \\
5714213232946368768  &  35.9 $\pm$ 4.9  & ~ 5714215741207424640  &  25.2 $\pm$ 3.7  \\
5714220620290111104  &  63.4 $\pm$ 5.2  & ~ 5714215225811379072  &   4.9 $\pm$ 2.3  \\
5714208525662494976  &  10.6 $\pm$ 6.5  & ~ 5714221857240628480  &  43.4 $\pm$ 5.5  \\
5714594351162455936  &  36.7 $\pm$ 7.9  & ~ 5714203474780825728  &   5.4 $\pm$ 4.5  \\
5714219864375836416  &  14.5 $\pm$ 6.2  & ~ 5714219245900567936  &  26.9 $\pm$ 5.2  \\
5714220379771938048  &  98.3 $\pm$ 5.7  & ~ 5714202787586066944  &  13.7 $\pm$ 3.0  \\
5714215908705903360  &  54.8 $\pm$ 6.3  & ~ 5714220517210900864  &   3.7 $\pm$ 3.7  \\
5714208525662501504  &  50.8 $\pm$ 7.5  & ~ 5714196121796947584  &  20.0 $\pm$ 5.4  \\
5714216084804886272  &  54.6 $\pm$ 4.2  & ~ 5714408804275744000  &  42.7 $\pm$ 4.2  \\
5714219280260304896  &  41.4 $\pm$ 4.7  & ~ 5714221444923778944  &  72.9 $\pm$ 5.4  \\
5714204917889795968  &  34.7 $\pm$ 7.5  &                       &                  \\
\hline
\end{tabular}
\end{center}
\end{table}

\begin{table*}
\caption{Chemical abundances derived with {\sf SYNTHE} ([X/H]) for giant stars observed with UVES in NGC\,2509. }\label{tab_abb_uves}
\begin{center}
\begin{tabular}{lcccccc}
\hline\hline
X & \scriptsize{5714216325322800128} & \scriptsize{5714215638128211968} & \scriptsize{5714216840718856064} & \scriptsize{5714215947365908352} & \scriptsize{5714218833583650304} & \scriptsize{5714209689591479424}  \\  
\hline                                                                      
Na   &  ~~0.28$\pm$0.10  &                   &  ~~0.31$\pm$0.10  &  ~~0.30$\pm$0.10  &  ~~0.36$\pm$0.10  &                   \\
Mg   & $-$0.10$\pm$0.15  &  ~~0.06$\pm$0.12  & $-$0.05$\pm$0.17  & $-$0.11$\pm$0.17  &  ~~0.03$\pm$0.16  & $-$0.06$\pm$0.08  \\
Al   &  ~~0.14$\pm$0.10  &  ~~0.09$\pm$0.07  &  ~~0.05$\pm$0.09  &  ~~0.12$\pm$0.10  &  ~~0.07$\pm$0.10  &  ~~0.14$\pm$0.11  \\
Si   &  ~~0.14$\pm$0.13  &  ~~0.12$\pm$0.10  &  ~~0.08$\pm$0.11  & $-$0.02$\pm$0.11  &  ~~0.13$\pm$0.08  & $-$0.03$\pm$0.12  \\
Ca   &  ~~0.11$\pm$0.06  &  ~~0.10$\pm$0.06  &  ~~0.06$\pm$0.07  &  ~~0.05$\pm$0.09  &  ~~0.16$\pm$0.14  &  ~~0.04$\pm$0.13  \\
Sc   &  ~~0.01$\pm$0.12  &  ~~0.03$\pm$0.13  &  ~~0.00$\pm$0.10  &  ~~0.02$\pm$0.12  &  ~~0.05$\pm$0.14  &  ~~0.10$\pm$0.12  \\
Ti   &  ~~0.06$\pm$0.10  &  ~~0.06$\pm$0.08  &  ~~0.03$\pm$0.08  &  ~~0.01$\pm$0.10  &  ~~0.03$\pm$0.10  & $-$0.02$\pm$0.10  \\  
V    &  ~~0.08$\pm$0.10  &  ~~0.14$\pm$0.11  &  ~~0.09$\pm$0.08  &  ~~0.03$\pm$0.10  &  ~~0.06$\pm$0.10  &  ~~0.08$\pm$0.11  \\
Cr   &  ~~0.14$\pm$0.12  &  ~~0.03$\pm$0.07  &  ~~0.07$\pm$0.12  &  ~~0.12$\pm$0.10  &  ~~0.14$\pm$0.11  &  ~~0.15$\pm$0.11  \\
Mn   &  ~~0.11$\pm$0.12  &  ~~0.12$\pm$0.11  &  ~~0.12$\pm$0.07  &  ~~0.10$\pm$0.10  &  ~~0.08$\pm$0.10  &  ~~0.03$\pm$0.12  \\
Fe   &  ~~0.17$\pm$0.07  &  ~~0.02$\pm$0.07  &  ~~0.03$\pm$0.09  &  ~~0.06$\pm$0.10  &  ~~0.16$\pm$0.11  &  ~~0.04$\pm$0.08  \\
Co   &  ~~0.24$\pm$0.14  &  ~~0.20$\pm$0.15  &  ~~0.18$\pm$0.19  &  ~~0.17$\pm$0.15  &  ~~0.13$\pm$0.14  &  ~~0.16$\pm$0.12  \\
Ni   &  ~~0.26$\pm$0.12  &  ~~0.10$\pm$0.14  &  ~~0.04$\pm$0.12  &  ~~0.08$\pm$0.15  &  ~~0.16$\pm$0.13  &  ~~0.10$\pm$0.12  \\
Zn   &                   &  ~~0.23$\pm$0.11  &  ~~0.25$\pm$0.10  &  ~~0.28$\pm$0.10  &  ~~0.30$\pm$0.09  &                   \\
Y    &  ~~0.22$\pm$0.08  &  ~~0.07$\pm$0.10  &  ~~0.12$\pm$0.12  &  ~~0.09$\pm$0.12  &  ~~0.10$\pm$0.11  &  ~~0.13$\pm$0.09  \\
Zr   &  ~~0.17$\pm$0.12  &  ~~0.14$\pm$0.15  &  ~~0.13$\pm$0.11  &  ~~0.13$\pm$0.11  &  ~~0.11$\pm$0.12  &  ~~0.14$\pm$0.12  \\
Ba   &  ~~0.24$\pm$0.10  &  ~~0.17$\pm$0.11  &  ~~0.12$\pm$0.10  &  ~~0.19$\pm$0.10  &  ~~0.21$\pm$0.12  &  ~~0.05$\pm$0.12  \\
La   &  ~~0.11$\pm$0.11  &  ~~0.08$\pm$0.11  &  ~~0.21$\pm$0.12  &  ~~0.29$\pm$0.11  &  ~~0.13$\pm$0.10  &  ~~0.16$\pm$0.09  \\
Ce   &  ~~0.25$\pm$0.10  &  ~~0.25$\pm$0.09  &  ~~0.31$\pm$0.11  &  ~~0.26$\pm$0.10  &  ~~0.29$\pm$0.10  &  ~~0.13$\pm$0.12  \\
Nd   &  ~~0.28$\pm$0.12  &  ~~0.14$\pm$0.11  &  ~~0.10$\pm$0.09  &  ~~0.14$\pm$0.09  &  ~~0.22$\pm$0.09  &  ~~0.11$\pm$0.09  \\
\hline
 & & & & & &  \\
 \hline
 X & \scriptsize{5714218283827858304} & \scriptsize{5714216737639647360} & \scriptsize{5714215191451661440}  & \scriptsize{5714216565840947328}  & \scriptsize{5714216698976454144} & \scriptsize{5714216119164382592} \\  
\hline
 Na  &                   &  ~~0.25$\pm$0.10  &                   &   ~~0.28$\pm$0.10  &                         &  ~~0.35$\pm$0.10  \\
 Mg  & $-$0.12$\pm$0.18  &                   &  ~~0.07$\pm$0.12  &  $-$0.09$\pm$0.18  &                         & $-$0.07$\pm$0.12  \\
 Al  &  ~~0.17$\pm$0.11  &  ~~0.07$\pm$0.10  &  ~~0.14$\pm$0.07  &   ~~0.13$\pm$0.08  &   ~~0.20$\pm$0.08  &  ~~0.09$\pm$0.11  \\
 Si  & $-$0.06$\pm$0.11  &  ~~0.00$\pm$0.12  &  ~~0.02$\pm$0.10  &   ~~0.12$\pm$0.07  &   ~~0.02$\pm$0.06  &  ~~0.06$\pm$0.11  \\
 Ca  &  ~~0.18$\pm$0.12  &  ~~0.04$\pm$0.09  &  ~~0.09$\pm$0.06  &   ~~0.12$\pm$0.12  &   ~~0.10$\pm$0.10  &  ~~0.05$\pm$0.07  \\
 Sc  &  ~~0.05$\pm$0.09  &  ~~0.08$\pm$0.13  &  ~~0.06$\pm$0.14  &   ~~0.19$\pm$0.09  &   ~~0.22$\pm$0.11  &  ~~0.13$\pm$0.13  \\
 Ti  &  ~~0.17$\pm$0.08  &  ~~0.10$\pm$0.09  &  ~~0.05$\pm$0.10  &   ~~0.06$\pm$0.09  &   ~~0.19$\pm$0.07  &  ~~0.16$\pm$0.08  \\  
 V   &  ~~0.16$\pm$0.08  &  ~~0.13$\pm$0.10  &  ~~0.09$\pm$0.10  &   ~~0.10$\pm$0.10  &   ~~0.19$\pm$0.10  &  ~~0.16$\pm$0.11  \\
 Cr  &  ~~0.12$\pm$0.13  &  ~~0.10$\pm$0.11  &  ~~0.03$\pm$0.08  &   ~~0.17$\pm$0.12  &   ~~0.20$\pm$0.09  &  ~~0.22$\pm$0.08  \\
 Mn  &  ~~0.19$\pm$0.12  &  ~~0.20$\pm$0.13  &  ~~0.17$\pm$0.11  &   ~~0.17$\pm$0.13  &   ~~0.17$\pm$0.12  &  ~~0.12$\pm$0.11  \\
 Fe  &  ~~0.17$\pm$0.09  &  ~~0.11$\pm$0.10  &  ~~0.11$\pm$0.11  &   ~~0.10$\pm$0.09  &   ~~0.18$\pm$0.08  &  ~~0.12$\pm$0.07  \\
 Co  &  ~~0.26$\pm$0.10  &  ~~0.21$\pm$0.19  &  ~~0.14$\pm$0.17  &   ~~0.22$\pm$0.09  &   ~~0.31$\pm$0.19  &  ~~0.29$\pm$0.15  \\
 Ni  &  ~~0.12$\pm$0.13  &  ~~0.17$\pm$0.12  &  ~~0.14$\pm$0.10  &   ~~0.18$\pm$0.08  &   ~~0.14$\pm$0.14  &  ~~0.11$\pm$0.09  \\
 Zn  &                   &                   &                   &   ~~0.33$\pm$0.10  &                         &                   \\
 Y   &  ~~0.17$\pm$0.10  &  ~~0.14$\pm$0.09  &  ~~0.07$\pm$0.11  &   ~~0.08$\pm$0.10  &        ~~0.16$\pm$0.08  &  ~~0.26$\pm$0.07  \\
 Zr  &                   &                   &  ~~0.08$\pm$0.10  &                    &    ~~0.20$\pm$0.11  &  ~~0.22$\pm$0.11  \\
 Ba  &                   &                   &  ~~0.15$\pm$0.10  &   ~~0.21$\pm$0.12  &                    &  ~~0.05$\pm$0.10  \\
 La  &                   &  ~~0.23$\pm$0.12  &  ~~0.12$\pm$0.12  &   ~~0.12$\pm$0.12  &   ~~0.27$\pm$0.13  &  ~~0.31$\pm$0.10  \\
 Ce  &  ~~0.10$\pm$0.11  &                   &  ~~0.11$\pm$0.11  &                    &   ~~0.32$\pm$0.11  &                    \\
 Nd  &  ~~0.21$\pm$0.11  &  ~~0.13$\pm$0.11  &  ~~0.15$\pm$0.12  &   ~~0.19$\pm$0.12  &   ~~0.26$\pm$0.10  &  ~~0.32$\pm$0.11  \\
\hline
\end{tabular}
\tablefoot{Solar references are taken from \citet{Grevesse07}.}
\end{center}
\label{tab_abb_uves}
\end{table*}

\begin{figure*}[]
\centering
\includegraphics[width=10cm]{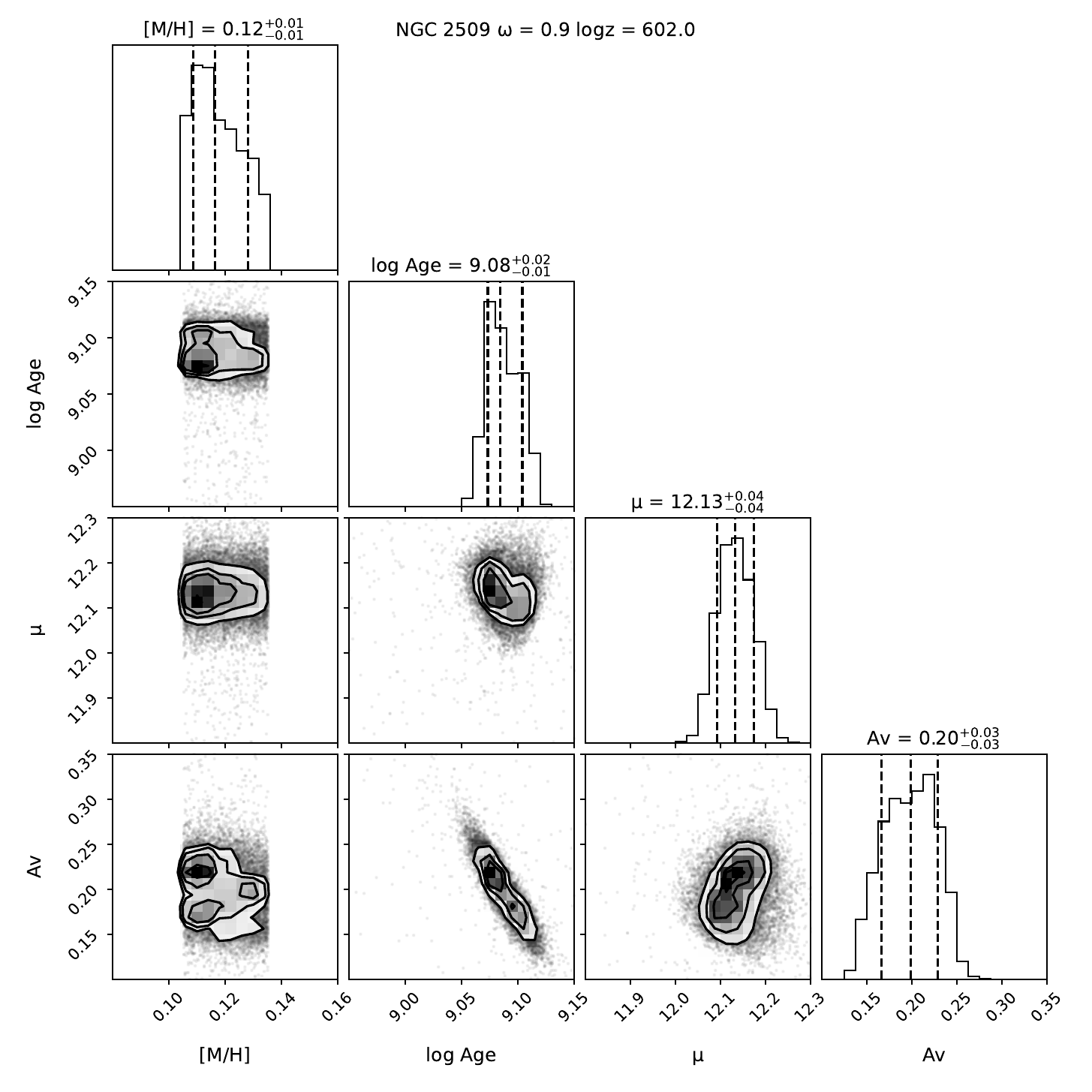}
\caption{Posterior distribution of the fitted parameters of NGC~2509 when $\Omega/\Omega_{crit}$=0.9, which is the best fit we obtained. The diagonal panels show the probability distribution functions of the variables with their medians and the 16th and 84th percentiles of the distribution (dashed lines). The maps show instead the 2D-probability
for each couple of parameters, highlighting possible correlations (e.g. between distance modulus and extinction or age and extinction).}
\label{CMD_solution}
\end{figure*}

\end{appendix}

\end{document}